**Article: Methods**

# Orthologs from maxmer sequence context

Kun Gao and Jonathan Miller

Physics and Biology Unit, Okinawa Institute of Science and Technology Graduate University, 1919-1 Tancha, Onna-son, Kunigami-gun, Okinawa, Japan 904-0495.

***Corresponding author:** Kun Gao (Email: kgao@oist.jp).

## Abstract

Context-dependent identification of orthologs customarily relies on conserved gene order or whole-genome sequence alignment. It is shown here that short-range context—as short as single maximal matches—also provides an effective means to identify orthologs within whole genomes. On pristine (un-repeatmasked) mammalian whole-genome assemblies we perform a genome "intersection" that in general consumes less than one thirtieth of the computation time required by commonly used methods for whole-genome alignment, and we extract "non-embedded maximal matches," maximal matches that are not embedded into other maximal matches, as potential orthologs. An ortholog identified via non-embedded maximal matches is analogous to a "positional ortholog" or a "primary ortholog" as defined in previous literature; such orthologs constitute homologs derived from the same direct ancestor whose ancestral positions in the genome are conserved. At the nucleotide level, non-embedded maximal matches recapitulate most exact matches identified by a Lastz net alignment. At the gene level, reciprocal best hits of genes containing non-embedded maximal matches recover one-to-one orthologs annotated by Ensembl Compara with high selectivity and high sensitivity; these reciprocal best hits additionally include putatively novel orthologs not found in Ensembl (e.g. over two thousand for human/chimpanzee). The method is especially suitable for genome-wide identification of orthologs.

# 1. Introduction

Sequences appearing in different genomes or within a single genome at frequencies beyond those expected on neutral evolution are expected to share common ancestors. Shared ancestry is known as *homology* and the corresponding genetic elements as *homologs* (Brown 2002). Homologs can be further classified as *orthologs* if they diverged via evolutionary speciation or *paralogs* if they diverged via duplication (Fitch 1970; Fitch 2000); see figure 1. Orthologs obtain special importance in terms of phylogeny (Fitch 2000; Blair and Hedges 2005; Ciccarelli et al. 2006; Altenhoff and Dessimoz 2012). It is generally believed that orthologs from genomes of different species often—if not always (Fang et al. 2010)—share similar function, while paralogs are more likely to develop new functions. Therefore, distinguishing between orthologs and paralogs is fundamental to the fields of comparative Omics, and is of great importance to our understanding of genome evolution and functional sequence innovation (Altenhoff and Dessimoz 2012).

Duplication and recombination events can make it difficult to ascertain the ancestry of genes in different organisms and thereby complicate the inference of orthologs and paralogs from modern-day genome sequences. Relationships among orthologs in two genomes are not necessarily one-to-one; because of different evolutionary histories, ortholog relationships exhibited in a phylogenetic tree can be *one-to-one*, or *one-to-many*, or *many-to-many*. Here *one* and *many* indicate the numbers of genes on each branch of the evolutionary tree bifurcating at a node that represents a speciation event. When further duplication occurs subsequent to the branching of two genomes (*recent* duplication relative to speciation), the corresponding paralogs are known as *in-paralogs*. One-to-many and many-to-many orthology arises from in-paralogs associated with the same ortholog through a speciation event; see figure 1 (b) and (c). Such a relationship is called *co-orthology* (Remm et al. 2001; Altenhoff and Dessimoz 2012). When a duplication event yielding a pair of paralogs in different genomes precedes the speciation of these two genomes (*ancient* duplication), the corresponding paralogs are called *out-paralogs* (or *between-species paralogs*); only when a pair of orthologs has undergone *recent* duplication in neither genome would the orthology between them remain one-to-one; see figure 1 (a).

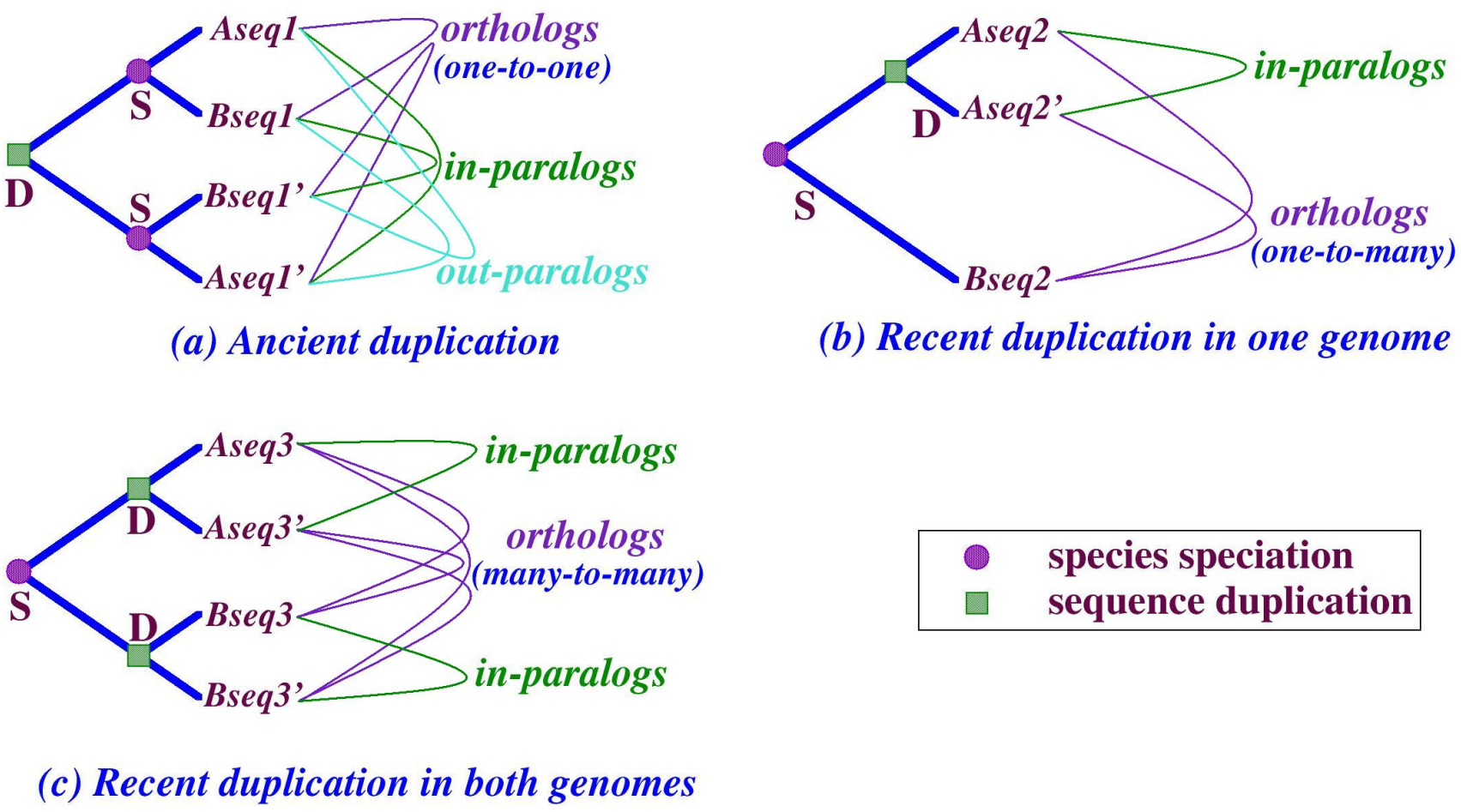

Figure 1. Customary definitions of orthologs and paralogs; see figure 1 of (Fitch 2000), and Ensembl documentation (http://www.ensembl.org/info/genome/compara/homology_method.html). In the figure, "***A***" and "***B***" denote two descendant genomes that diverged via a speciation event; ***seq1/seq1'*** (***seq2/seq2'***, ***seq3/seq3'***) indicate the parent and daughter copies of a gene that diverged via a duplication event in genome ***A*** or ***B***.

Customarily, orthologs can be identified through two categories of methods: *tree-based* methods and *graph-based* methods; some other methods represent hybrids thereof (Kuzniar et al. 2008).

Tree-based methods infer orthology and paralogy from phylogenetic trees; they provide phylogenetic resolution as in figure 1 at multiple levels of a gene tree. Phylogenetic analysis of the gene lineage is thought in principle to enable the strongest discrimination between orthology and paralogy; however, such an approach is computationally intensive for large datasets and difficult to automate. Graph-based methods evaluate one pair of genomes at a time; inference relies on pairwise all-versus-all sequence comparison. Such methods are less computationally demanding and more suitable for inferring orthology within two or more complete genomes.

In lieu of phylogenetic inference, graph-based methods usually "rely on some sort of shortcut, or working definition, to detect orthology" (Moreno-Hagelsieb and Latimer 2008; Ward and Moreno-Hagelsieb 2014). A standard graph-based method argues that within a given genome pair, orthologs are most likely to be those homologs that diverged least (Altenhoff and Dessimoz 2012); a well known working definition of orthology is based upon reciprocal/bidirectional best hits (RBHs or BBHs for short) (Tatusov et al. 1997; Bork et al. 1998; Overbeek et al. 1999; Wolf and Koonin 2012). Therefore, graph-based methods are most typically predisposed to yield one-to-one (or nearly one-to-one) ortholog relationships. Such one-to-one ortholog relationships include not only the one-to-one orthologs in figure 1 (a) but also some of the co-orthologs in figure 1 (b) and (c). For instance, figure 1 (b) in (Dalquen and Dessimoz 2013) shows an example in which a group of co-orthologs between human and mouse genomes is refined into a "BBH and orthologs" pair and a "non-BBH and orthologs" pair; a standard BBH approach will identify the former as a one-to-one ortholog relationship. Thus, a proclivity for one-to-one ortholog relationships is implicitly built into the "graph construction phase" of many graph-based methods, although some of them, such as the well-known *Inparanoid* (Remm et al. 2001), subsequently extend these one-to-one orthologs to encompass co-orthologs in a "clustering phase" that fits phylogenetic inferences.

One-to-one orthologs inferred by graph-based methods can be expected to represent the "most orthologous" pairs of genes between the compared genomes; they are more likely to play equivalent roles within both genomes than are other homologs. Over the past decade, different research groups have applied a variety of terms to describe these orthologs, among them *true exemplar* (Sankoff 1999), *positional ortholog* (Koski et al. 2001; Swidan et al. 2006), *main ortholog* (Remm et al. 2001; Fu et al. 2007), *super ortholog* (Zmasek and Eddy 2002) and *true ortholog* (Bandyopadhyay et al. 2006). Although the definitions of these terms remain incomplete, operational, and are not fully consistent with one another (Dewey 2011), this nomenclature distills two essential features of these orthologs: (1) such orthologs most faithfully reflect the original positions of their ancestral sequences in the common ancestor's genome; and (2) such orthologs are directly inherited from the ancestral sequences and are not the products of recent duplications. Based on previous studies, two conceptual definitions were proposed: *primary ortholog* by Han and Hahn (Han and Hahn 2009), and a redefined *positional ortholog* by Dewey (Dewey 2011). These two definitions revealed a key to identifying such orthologs: when a *recent* duplication occurs, primary/positional orthology only applies to its *original* (or *parent*) copy, but not to the *derived* (or *daughter*) copies.

In order to discriminate among co-orthologs and identify "most orthologous" pairs of genes, we must therefore distinguish between the parent and daughter copies of in-paralogs. In-paralogs that branched at around the same time from their orthologous cognate are expected to exhibit comparable sequence divergence from their orthologous cognate, so that sequence similarity alone is often not sufficient to distinguish co-orthologs from one another (Dewey 2011). Nevertheless, genes at different locations are also likely to exhibit differences in their *neighborhoods* within the genome: parent and daughter copies of a duplication, while similar to one another in sequence content, are often embedded into different *genomic contexts* that did *not* themselves also undergo duplication at the same time, if at all. Here the general notion of "genomic context" provides a tool to differentiate the "most orthologous" pairs of genes from all other homolog pairs. Han *et al.* observe that "*[W]hen compared to an outgroup gene that has the same ancestor, the parent copy is expected to maintain longer stretches of conserved synteny within its flanking region, while the daughter copies exhibit only shorter syntenic blocks that comprise the duplicated segments*" (Han et al. 2009); similar idea has been incorporated into ortholog identification in previous literature. Most commonly, synteny information—in the sense of "conserved gene order"—is taken as an indicator of genomic context and incorporated into methods that are primarily based on sequence similarity or on gene evolution models (Fu et al. 2007; Dewey 2011).

In this paper, (1) we propose a conceptual definition for the "most orthologous" pairs of genes or sequences between two genomes; our definition is fundamentally consistent with, but slightly different from, Han and Hahn's "primary orthologs" and Dewey's "positional orthologs." Based solely on evolutionary history—the full evolutionary path connecting descendants to their most recent common ancestor—our definition can be explicitly visualized on gene trees as in figure 1 (see figure 3 below). As a generalization of the basic idea of invoking genomic context to distinguish between parent and daughter copies of duplication, (2) we suggest a new graph-based method to infer these orthologs from whole genome nucleotide sequences. We distinguish between parent and daughter copies of a duplication (and therefore among different classes of homologs) not as did Han *et al.* solely by *length* of conserved synteny, but rather by a sequence-based property of "*embedding*" among context-dependent maximal matches: *homologs that are embedded in other homologs are thought of as "less orthologous" than the homolog pair into which they are embedded* (see details in section 2.3)—as exhibited in figure 2. In practice, our method infers orthology and paralogy genome-wide by exploiting *local contexts of matches* at scales much shorter than a gene, whereas all previous methods rely on conserved synteny or large regions of colinearity (Dewey 2011) over relatively long domains that may well span a series of genes (this is obvious by comparing figure 2 to the figure 2 of (Han and Hahn 2009) for example).

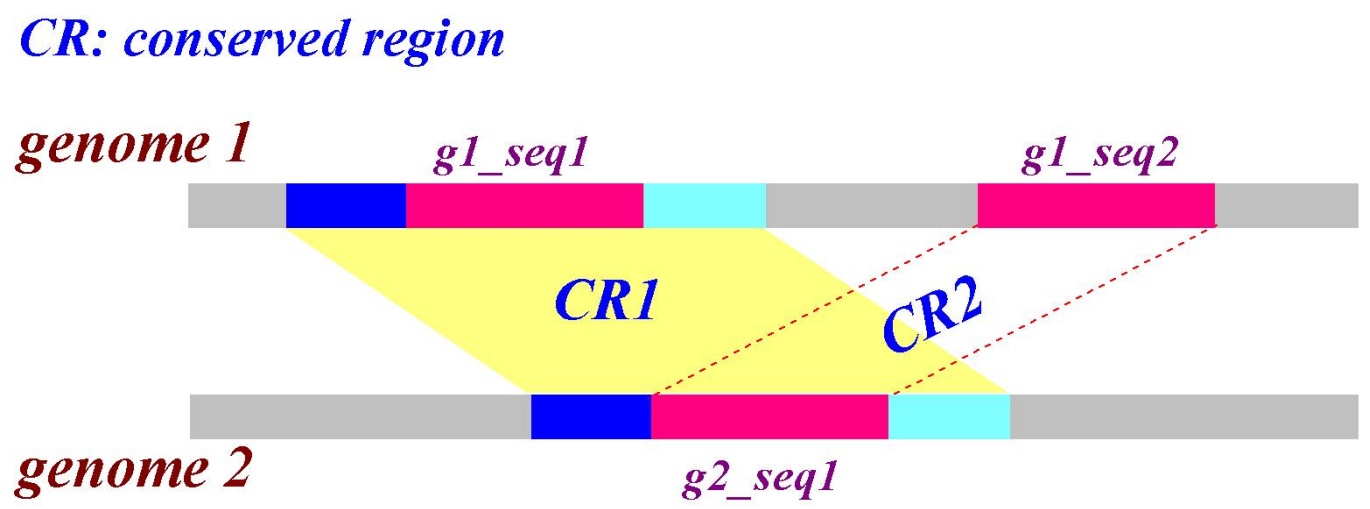

Figure 2. Schematic of how we propose to identify "most orthologous" pairs of sequences between two genomes. Apart from grey, identical colors indicate highly conserved regions: ***g1_seq1*** and ***g1_seq2*** respectively are parent and daughter copies of a duplication in ***genome 1***; ***g2_seq1*** is their common homologous cognate in ***genome 2***. To Han *et al.*'s viewpoint, the extent of "synteny" conserved between ***g1_seq1*** and the outgroup sequence ***g2_seq1*** is evidently greater than that between ***g1_seq2*** and ***g2_seq1***—indeed, conserved region ***CR2*** is "embedded" into conserved region ***CR1***—therefore ***g2_seq1*** is thought of as "more orthologous" to ***g1_seq1*** than to ***g1_seq2***. This embedding property is recapitulated at the nucleotide level by context-sensitive maximal matches (see section 2.2).

The paper is organized as following: in section 2, we define *sequence orthology* and propose a new context-based method to identify sequence orthologs for a pair of genomes; we propose and demonstrate the relationship to sequence orthology of a sequence-based property of "*embedding*" among maximal matches. In section 3, we show the effectiveness of our method: in section 3.1, we apply a series of numerical simulations to test our method under different models of genome evolution, while in section 3.2 and 3.3, we compare orthologs in natural genomes inferred by our method with those inferred by a Lastz net alignment on the nucleotide level, and with those annotated by Ensembl Compara on the gene level.

# 2. Definitions and New Approaches

## 2.1 *Lineal orthologs and collateral orthologs*

The notion of sequence homology extends naturally from protein-coding genes to genomic sequences in general. A genomic sequence is characterized by its constituent string of bases and by its position in the genome; two genomic sequences are thought of as "the same sequence" only if they occupy the same site in a genome (i.e., they share not only the same string but also the same coordinates). We designate a pair of orthologous sequences as ***lineal orthologs*** if both sequences are "directly inherited" from the same ancestral sequence of their most recent common ancestor's genome, where "directly inherited" means that vertical heredity (inheritance from parent to offspring) is the only evolutionary process connecting ancestor to descendant; see blue edges in figure 3. In the context of direct inheritance, we call the ancestor and descendant respectively the *direct ancestor* and *direct descendant* of each other. Lineal orthologs are orthologs that have the same direct ancestor; we designate orthologs having distinct direct ancestors as ***collateral orthologs***[1].

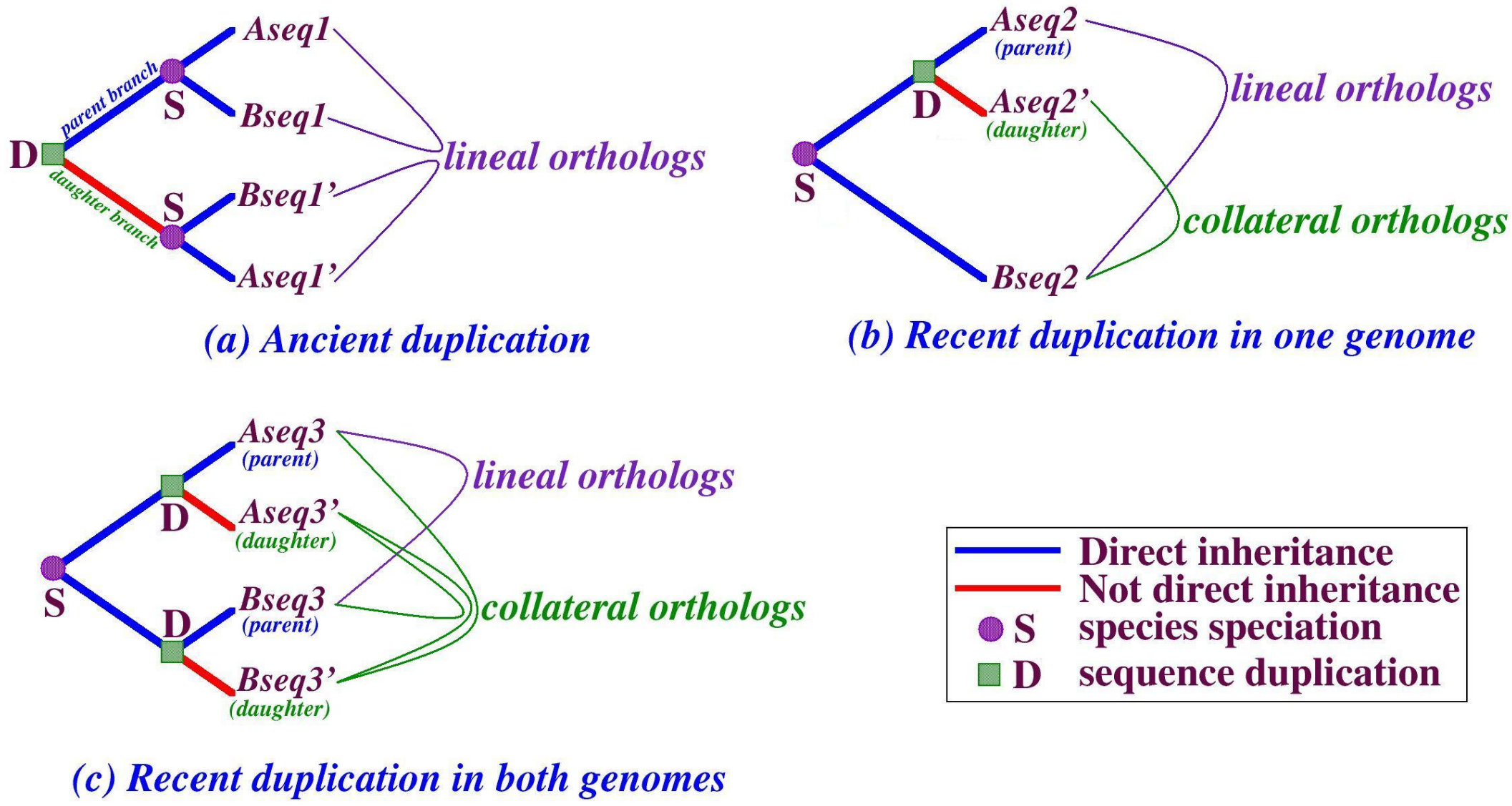

Figure 3. Definitions of lineal orthologs and collateral orthologs. Subfigures (a)–(c) show exactly the same evolutionary histories as subfigures (a)–(c) of figure 1. Blue edges indicate evolutionary paths along which direct inheritance applies, while red edges indicate creation of new duplicates; the evolutionary path connecting a pair of lineal orthologs must consist of blue edges only.

We have defined lineal and collateral orthology in terms of the full evolutionary path connecting ancestor to descendant—not only the evolutionary event through which the orthologs first diverge, but also the evolution of each lineage subsequent to branching. Our definition of lineal orthology permits duplication events within these lineages; nevertheless, direct inheritance applies solely to the parent copy: only the parent copy of a duplication can be a direct descendant of the ancestor. Daughter copies, on the other hand, are considered as products newly created by the process of duplication—they can thenceforth have their own direct descendants (see ***Aseq1'*** and ***Bseq1'*** in figure 3 (a)), but they don't have a direct ancestor. In the customary definition of ortholog (Fitch 2000) (see figure 1), daughter copies of a recent duplication are thought of as having diverged from their orthologous cognate through a speciation event; from our perspective the customary definition is troublesome because these daughter copies had not yet been created at the time of the speciation event. In contrast, according to our definition, daughter copies of a recent duplication contribute only to collateral orthologs. Figure 3 (a)–(c) illustrate our definitions of lineal orthologs and collateral orthologs under exactly the same evolutionary histories as in subfigures (a)–(c) of figure 1. By our definition, one-to-one orthologs in figure 1 (a) are always lineal orthologs, but one-to-many and many-to-many orthologs in figure 1 (b) and (c) can also be elucidated: if neither sequence of a pair of orthologs is the daughter copy of a recent duplication—so that the full evolutionary paths connecting these orthologs to their latest common ancestor represent direct inheritance—then the pair of orthologs is lineal; if either sequence of a pair of orthologs is the daughter copy of a recent duplication, then the pair of orthologs is collateral; this distinction is consistent with figure 1 (b) of (Dalquen and Dessimoz 2013).

[1] Our usage of the terms "lineal" and "collateral" derive from property law: "a lineal descendant is one in the direct line of descent, while a collateral descendant is one descended from the same ancestor but not in the same line, e.g. brothers and sisters" (see Brake, M.E., *The Beginnings of Property Law—Part III: The Evolution of Conveyancing*, 16 U. Det. L.J. 61 (1952-1953), at page 71 (footnote).).

Lineal orthology strictly represents a one-to-one relationship; in contrast to "orthology," lineal orthology is transitive: if sequences $A$ and $B$ are a pair of lineal orthologs, and sequences $B$ and $C$ another pair of lineal orthologs, then sequences $A$ and $C$ are also a pair of lineal orthologs. Therefore, within a group of lineal orthologs, every sequence pair comprises a pair of lineal orthologs. As in earlier work (Sankoff 1999; Koski et al. 2001; Remm et al. 2001; Bandyopadhyay et al. 2006; Fu et al. 2007; Han and Hahn 2009; Dewey 2011), we assume that the positions of lineal orthologs within present-day genomes faithfully reflect the original positions of their direct ancestors; this assumption will be applied to distinguish lineal orthologs from other homologs.

## 2.2 *Non-embedded and embedded maximal matches*

We define a *maximal match* between two genomes as a contiguous run of matching bases—subject to specified matching criteria—that is extendable at neither end, represented as a *set* of two highly similar (or identical) strings, one in each genome. Two maximal matches are said to overlap with each other if either sequence of one maximal match shares a site with a sequence of the other maximal match. Depending on the types of overlaps among their constituent sequences, overlaps between two maximal matches fall into the following three categories (see figure 4):

(1) *Embedded overlap*: one maximal match is said to be *embedded* in another if the sites spanned by either sequence of the former are *properly* contained in the sites spanned by the corresponding sequence of the latter; see figure 4 (a).

(2) *Full overlap*: two maximal matches constitute a "full overlap" if they span exactly the same sites in one of the genomes, and their sequences in the other genome are not embedded in each other; see figure 4 (b).

(3) *Partial overlap*: two maximal matches constitute a "partial overlap" if their span includes common sites in either or both genomes, but satisfies neither (1) nor (2); see figure 4 (c).

Based on these three types of overlaps, all maximal matches between two genomes can similarly be classified into two mutually exclusive groups: if a maximal match is embedded in another maximal match, it is classified as an ***embedded maximal match***—red bars in figure 4 (a) represent embedded maximal matches; if a maximal match is not embedded in *any other* maximal match, it is a ***non-embedded maximal match***.

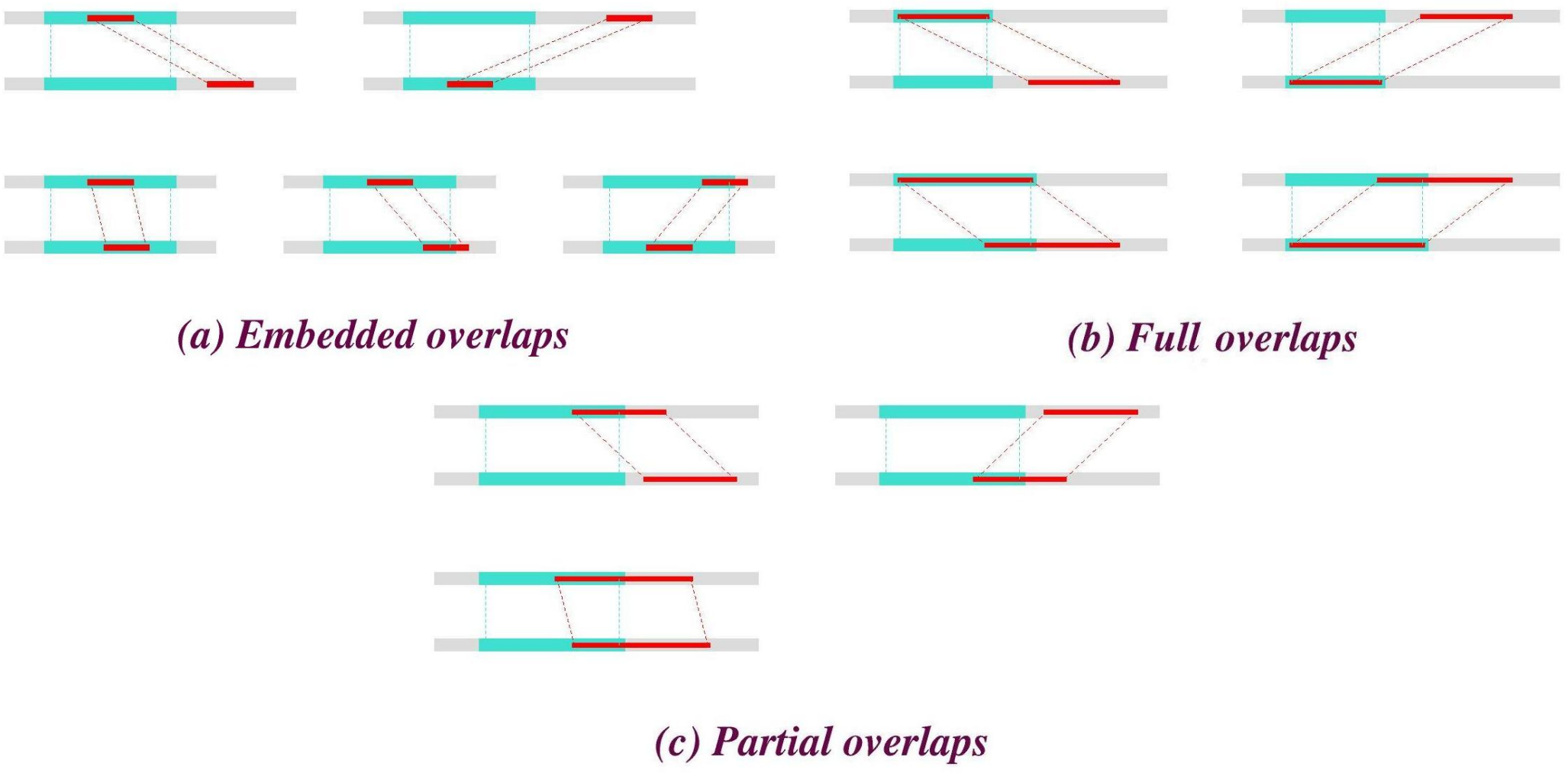

Figure 4. Types of overlaps between maximal matches. Grey bars represent the compared genomes. Red bars and turquoise bars represent two different maximal matches; these two maximal matches are said to overlap with each other if either of their sequences overlap.

## 2.3 *Identifying lineal orthologs by non-embedded maximal matches*

We anticipate that lineal orthologs can be distinguished from collateral orthologs and paralogs by means of their associated non-embedded and embedded maximal matches respectively. We demonstrate that collateral orthologous and paralogous regions are often embedded in lineal orthologous regions (see figure 5), so that ***embedded maximal matches are associated primarily with collateral orthologs and paralogs, whereas non-embedded maximal matches are associated primarily with lineal orthologs.***

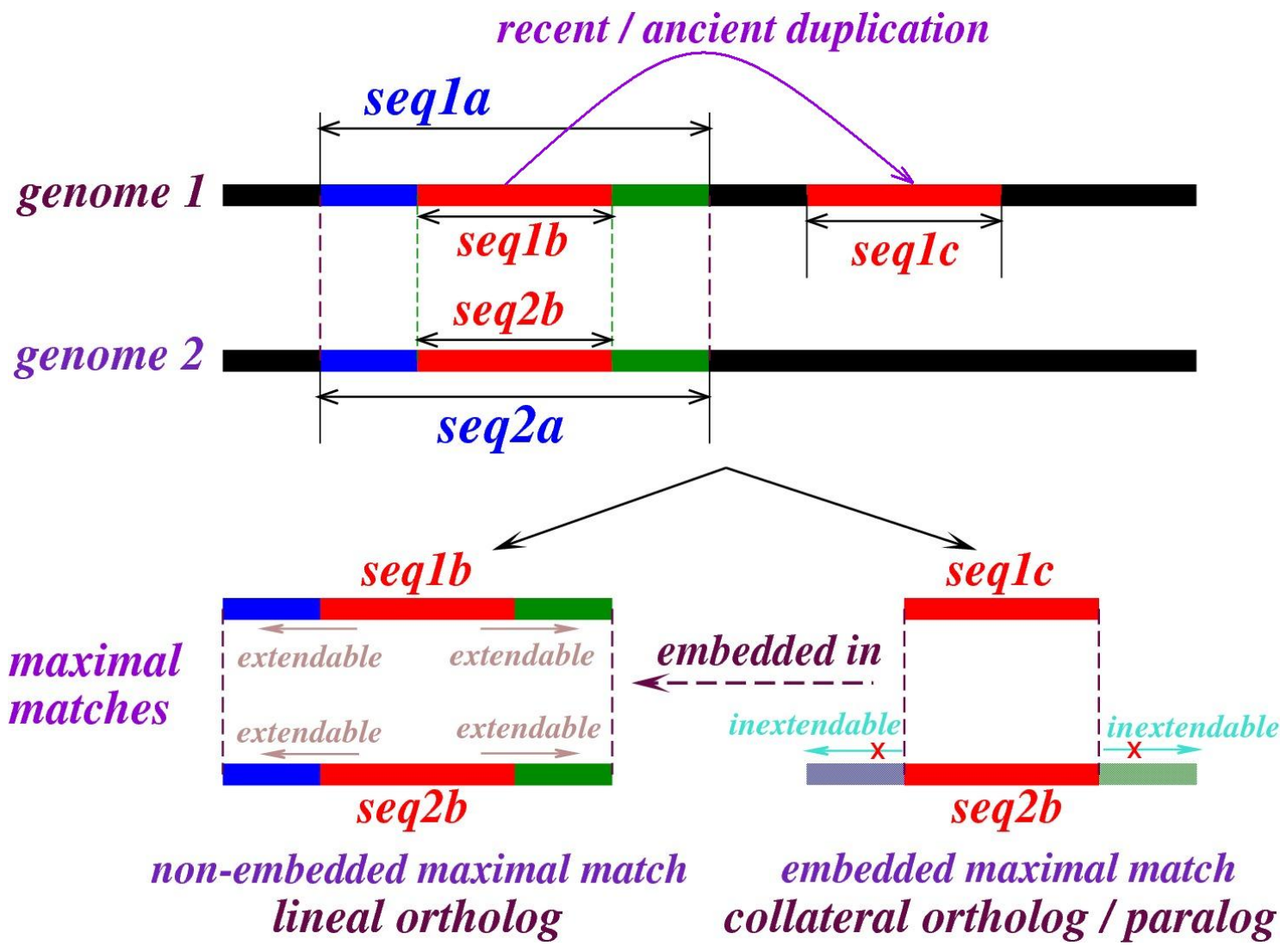

Figure 5. Collateral orthologous and paralogous regions tend to be embedded in lineal orthologous regions. Apart from black, bars of the same color indicate highly conserved genomic regions. The solid arrow indicates the direction of a segmental duplication from ***seq1b*** to ***seq1c***, which can be either recent or ancient. When comparing with an outgroup sequence ***seq2b*** in a different genome, the match between the parent copy ***seq1b*** and ***seq2b*** can usually be extended into the flanking region, whereas the match between the daughter copy ***seq1c*** and ***seq2b*** is not extendable. As a result, the latter match is embedded in the former.

Figure 5 illustrates the above proposal by showing a representative example of a collateral orthologous/paralogous region embedded in a lineal orthologous region. ***seq1b*** and ***seq1c*** in ***genome 1*** are respectively the parent and daughter copies of a segmental duplication; relative to the speciation from ***genome 2***, this segmental duplication could be either recent or ancient. ***seq2b*** in ***genome 2*** is the lineal orthologous counterpart of ***seq1b***. Therefore, in the comparison of these two genomes, ***seq2b*** can match both ***seq1b*** and ***seq1c***: the match between ***seq1b*** and ***seq2b*** constitutes a pair of lineal orthologs, whereas the match between ***seq1c*** and ***seq2b*** constitutes a pair of collateral orthologs if the duplication from ***seq1b*** to ***seq1c*** is recent, or a pair of paralogs if that duplication is ancient.

Next we account for the flanking regions of these sequences. As a pair of lineal orthologs, ***seq1b*** and ***seq2b*** are expected to share the same genomic context as their most recent common ancestor, whereas ***seq1c*** is expected to exhibit a different context. As a result, the match between ***seq1b*** and ***seq2b*** is probably not maximal: it most likely extends into the flanking regions (blue and green regions in figure 5) until local sequence variations terminate it on both ends. In contrast, differing contexts prevent the match between ***seq1c*** and ***seq2b*** from extending into flanking regions; in most cases this match is constituted solely by the duplicated region (red regions in figure 5), so that the maximal match between ***seq1c*** and ***seq2b*** is embedded into that between ***seq1a*** and ***seq2a***. Therefore, we expect most collateral orthologs and paralogs to be embedded, and lineal orthologs correspondingly non-embedded.

In the above discussion, we neglected the impact of sequence variation within the regions of duplication. In the ideal or nearly ideal case that all homologous regions of interest are well conserved, the above interpretation accounts for the effectiveness of our method. On the other hand, when sequence variation is frequent, recent variations in the parent copy of duplication may degrade the lineal orthologous match and yield misclassification of orthologs and paralogs (see supplementary figure 1 for some representative examples). However, it is generally believed that orthologs that have conserved their ancestral genomic positions are under greater evolutionary constraint than other homologs (Koski et al. 2001; Notebaart et al. 2005; Burgetz et al. 2006; Cusack and Wolfe 2007; Lemoine et al. 2007; Jun et al. 2009b; Wang et al. 2010; Dewey 2011), whereas duplicates in non-ancestral positions are more likely to undergo positive selection (Han et al. 2009). Therefore, misidentification exhibited in supplementary figure 1 could be expected to be minor.

Our proposal for *sequences* parallels the one for *genes* by Han and Hahn, who claim that the parental duplicate shares greater synteny with an outgroup gene than do its daughter duplicates (Han and Hahn 2009); a difference—and it is technically and practically a very big difference—is rather than working with genes we work directly with nucleotide matches. Since the lengths of maximal matches between two vertebrate genomes are typically between 20 and 3000 nucleotide bases while genes of the same species span as much as 2~3 Mbases, our method is, in a certain fundamental sense that we explain below, "local" rather than "global." We can apply our method directly to whole genome sequences, irrespective of genes; it enables genome-wide discrimination of lineal orthologs from collateral orthologs and paralogs from the genomic contexts of single maximal matches.

To obtain all non-embedded and embedded maximal matches between two genomes, we must first of all specify the matching criteria and compare their sequences either by intersection or by alignment. In this paper, our non-embedded and embedded maximal matches are obtained without loss of generality from 4-base exact matched genome intersections on pristine (un-repeatmasked) whole-genome sequences (see section 5.2 and supplementary material 1 for computation details);

non-embedded and embedded maximal matches identified in a Lastz raw alignment are referred to as "*non-nested and nested CMRs*" in (Gao and Miller 2014). Alternatively, we also examine another set of maximal matches—maximal unique matches ("*MUMs*" for short) (Delcher et al. 1999; Delcher et al. 2002; Kurtz et al. 2004)—as an approximation to the non-embedded maximal matches defined in this paper. *MUMs* are required to be unique in both compared genomes; they form a proper subset of all non-embedded maximal matches (see section 5.2), but permit no "full overlaps" (see figure 4 (b)). For most genomes, *MUMs* recover a large majority of all non-embedded maximal matches, and yield qualitatively similar outcomes for the calculations described in this paper. However, under certain evolutionary scenarios, full overlaps involve not only collateral orthologs or paralogs, but also lineal orthologs (see supplementary figure 1 (c) for example). Therefore, ignoring all full overlaps and identifying lineal orthologs by *MUMs* only would apparently result in a higher selectivity (i.e., a higher true positive rate), but at the same time a lower sensitivity (i.e., a higher false negative rate).

# 3. Results

## 3.1 *Numerical simulation with a genome growth model*

To evaluate the effectiveness of the method introduced in section 2.3 under certain models of genome evolution, we implement a series of numerical simulations with a "genome growth model" that is characterized by two dynamical elements only: random segmental duplication and random point substitution; obviously, evolutionary selection is absent from these dynamics. This model was earlier proposed by Chen *et al.*; by "growth model" the authors referred to "a computer algorithm for generating, from an initial sequence, a target sequence that has a given profile and other specific genome-like attributes" (Chen et al. 2010). In 2013, Massip and Arndt (Massip and Arndt 2013; see also Koroteev and Miller, unpublished data, http://arxiv.org/abs/1304.1409v3, last accessed October 29, 2013) reported that such a model yields repetitive sequences within a single genome whose length distribution at stationarity resembles those observed in natural genomes (Gao and Miller 2011; see also Koroteev and Miller 2011). In this section, we implement this model to simulate genome sequence evolution: within each time unit, we introduce $m$ random segmental duplications and $n$ random point substitutions to the genome; each segmental duplication substitutes a segment of consecutive bases of length $K$=1000 at a random position in the genome with a segment of the same length copied from another random position, while each point substitution randomly alters one nucleotide. Such an evolutionary model does not change the size of the genome; the length distribution of duplicates in stationary state depends only on the $m/n$ ratio (Massip and Arndt 2013).

Starting from a 4-base (A, T, G, C) random sequence, we simulate the genome growth dynamics until the duplication length distribution stabilizes. We take a sequence from this stationary state as the common ancestor's genome, and study its subsequent divergence into two lineages: for each lineage, we simulate subsequent evolution by continuing the growth dynamics independently. During this process, we record the positions of all recent duplications and substitutions in these lineages so that we can distinguish among lineal orthologs, collateral orthologs and paralogs between the mutated genomes. We compute a whole-genome intersection of the two mutated genomes, extract all embedded (*em* for short) and non-embedded (*nem* for short) maximal matches that are no shorter than 20 consecutive bases (see section 5.2 for computation details), and investigate their respective consistency with lineal orthologs (*lo* for short), and collateral orthologs plus paralogs (*cop* for short). For that purpose, we study four conditional probabilities: P(*lo*|*nem*), P(*cop*|*em*), P(*nem*|*lo*) and P(*em*|*cop*) among these four sets of maximal matches (see figure 6), in

which the former two probabilities indicate the selectivity of the method, whereas the latter two probabilities indicate the sensitivity of the method. The conditional probabilities P($A|B$) for sets of maximal matches $A$ and $B$ represents the probability of a maximal match belonging to $A$ given that it belongs to $B$; for example, the probability that a non-embedded maximal match is constituted by lineal ortholog sequences is:

$$\mathrm{P}(lo|nem) = \frac{\#\left(\textit{non-embedded maximal matches that are lineal orthologs}\right)}{\#\left(\textit{all non-embedded maximal matches}\right)},$$

where #() represents the total number of bases contained in the indicated set of maximal matches. Unless otherwise indicated, all probabilities are weighted by maximal match length in bases; using unweighted probabilities instead yields only minor differences.

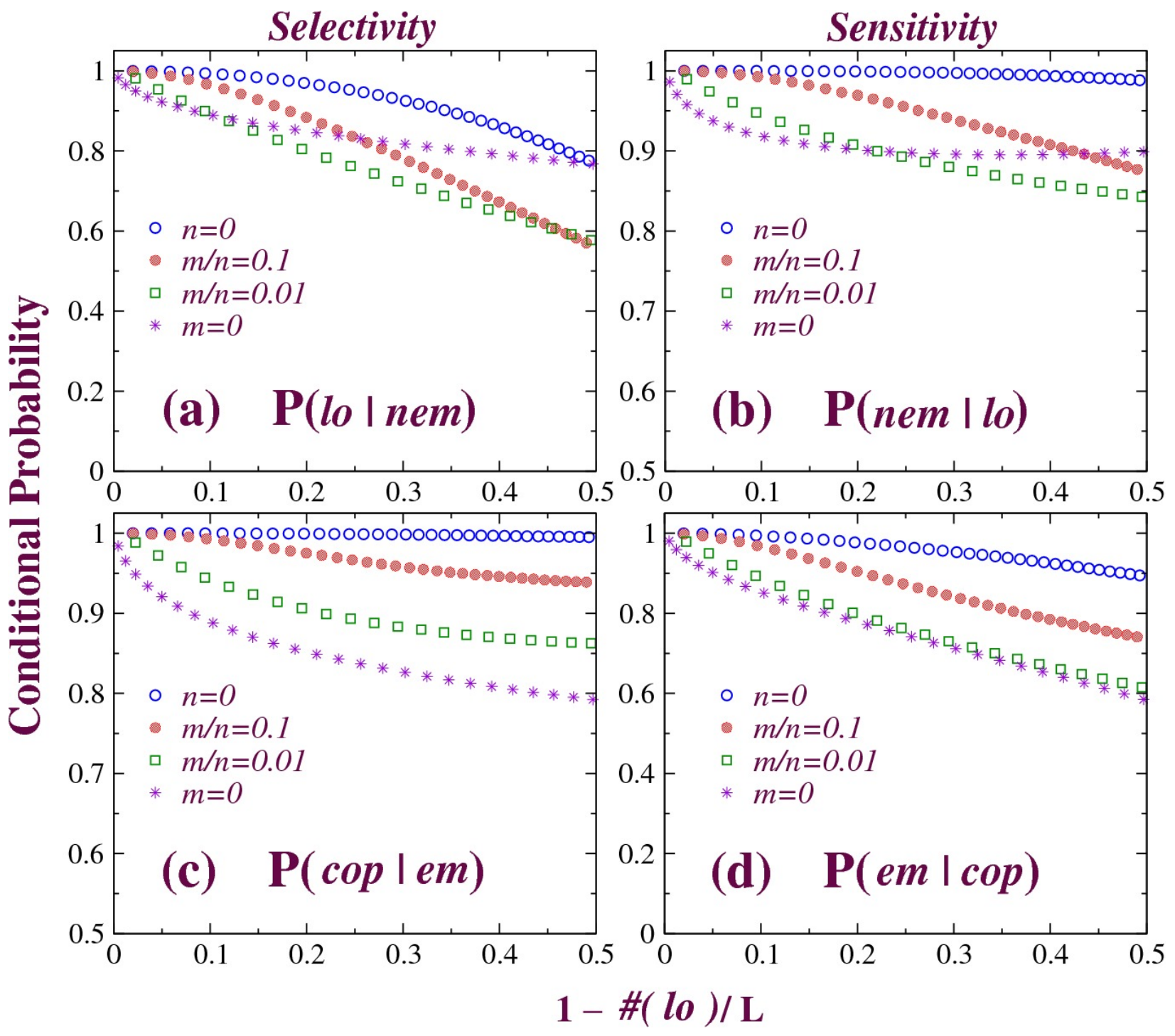

Figure 6. Conditional probabilities derived from numerical simulations that a pair of matched sequences are (a) lineal orthologs ("*lo*"), given that they belong to a non-embedded maximal match ("*nem*"); (b) belong to a non-embedded maximal match given that they are lineal orthologs; (c) collateral orthologs or paralogs ("*cop"*) given that they belong to an embedded maximal match ("*em*"); and (d) belong to an embedded maximal match given that they are collateral orthologs or paralogs; for all subfigures, probabilities are calculated over all maximal matches no shorter than 20 nucleotide bases, and the x-axis shows the proportion of genomes that are *not* covered by lineal orthologs ($\geqslant$20 bases), as an indication of the evolutionary distance between the mutated genomes. The common ancestor's genome was generated from a 4-base random sequence of length $L=10^8$ nucleotide bases by a genome growth model (Chen et al. 2010) with parameters $m/n$ = 0.01 (Massip and Arndt 2013); starting from this common ancestor's genome, two subsequent lineages evolve independently: within each time unit, $m$ random segmental duplications of length $K$=1000 nucleotides and $n$ random point substitutions are introduced to each lineage.

In principle, the performance of our method depends on certain parameters of the model of evolution (e.g. $m/n$); however, detailed characterization of the parameter space of the model exceeds the scope of this paper. In figure 6, we simulate the evolution of genomes after their divergence from each other with four representative parameter values: $n=0$ (duplication only), $m/n=0.1$ (high duplication-substitution rate), $m/n=0.01$ (low duplication-substitution rate) and $m=0$ (substitution only). We use the complement of the coverage of lineal orthologs (no shorter than 20 bases) in the mutated genomes to describe their evolutionary distance. It can be seen in figure 6 that when the evolutionary distance between the mutated genomes is not too far—in the sense that most of the lineal orthologs between these genomes remain well conserved—then both the selectivity and the sensitivity of our method remain high.

Figure 6 also reflects two characteristics of this method: P(*cop*|*em*)>P(*lo*|*nem*), reflecting that it is more selective in identifying *cop* than identifying *lo*; and P(*nem*|*lo*)>P(*lo*|*nem*), reflecting that it has a higher sensitivity than selectivity in identifying *lo*. Both these characteristics can be attributed to the fact that it requires more variations in the genome to misidentify *lo* as *em* than to misidentify *cop* as *nem*—which is obvious in supplementary figure 1 (a) and (b).

*MUMs* recover a large majority of all *nem*, thus they can also be taken as candidates for lineal orthologs. Supplementary figure 2 shows a comparison between the performance of *MUMs* and *nem* in the above numerical simulations; as we discussed in section 2.3, *MUMs* exhibit a higher selectivity (*P*(*lo*|*MUMs*)>*P*(*lo*|*nem*)), but a lower sensitivity (1-*P*(*MUMs*|*lo*)>1-*P*(*nem*|*lo*)) than *nem*. On the other hand, lineal orthologs by definition have no overlaps with one another: full or partial overlaps among *nem* often emerge as byproducts of misidentification of lineal orthologs (see supplementary figure 1); this enables us to estimate an upper bound on the false positive rate in our lineal ortholog identification from the ratio of overlaps among *nem* (see supplementary material 2).

### 3.2 *Comparison on nucleotide level with a Lastz net alignment*

As an example in real genomes, we compare the lineal orthologs identified by our method with those extracted from a whole-genome *alignment*. Whole-genome alignment is an independent method of ortholog identification, which exploits synteny information for chaining and netting; for example the Blastz/Lastz net alignment concatenates single matches into chains and purge repetitive hits, with only the best-matched hits returned. Kent *et al*. claim that Blastz net alignment may discriminate between orthologs and paralogs (Kent et al. 2003); however, from the perspective of this paper, orthologs encompassed within such alignment—especially within the "single best chain"—are expected to be lineal orthologs.

Table 1. Comparison between non-embedded maximal matches ("*nem*") and maximal matches in a Lastz net alignment ("*net*").

| Genome Pairs | Conditional Probabilities | Forward Strand | | Reverse Strand | |
|---|---|---|---|---|---|
| | | Whole Genome | NRM Regions[a] | Whole Genome | NRM Regions[a] |
| Human *vs.* Chimp | *P*(*net*\|*nem*) | 87.9% (2158550540/2454970619) | 97.8% (1047313326/1070330999) | 46.9% (213325366/455298641) | 88.9% (101183582/113820125) |
| | *P*(*nem*\|*net*) | 96.4% (2158550540/2238382261) | 98.7% (1047313326/1061387575) | 91.5% (213325366/233109466) | 94.0% (101183582/107657799) |
| Human *vs.* Mouse | *P*(*net*\|*nem*) | 12.9% (6559742/50913232) | 92.0% (6123123/6653203) | 11.2% (5540511/49607699) | 90.5% (5161741/5706103) |
| | *P*(*nem*\|*net*) | 97.1% (6559742/6752332) | 97.7% (6123123/6268416) | 96.6% (5540511/5735168) | 97.2% (5161741/5311032) |

Note— Percentages show the conditional probabilities that a maximal match is in one set given that it is in the other; numbers in brackets exhibit the total number of nucleotides contained in the corresponding sets of maximal matches.

[a] NRM Regions: genome regions that are not repeat-masked.

We compare our non-embedded maximal matches identified by intersection (*nem* for short, see section 5.2 and supplementary material 1 for computation details) with exact matches returned by Lastz net alignment (*net* for short, see (Gao and Miller 2011) and (Gao and Miller 2014) for computation details). Table 1 shows such comparisons for two pairs of genomes: human/chimp and human/mouse. To remove the effect of random matches at short length scales (Salerno et al.

2006; Koroteev and Miller 2011; Taillefer and Miller 2011a; Massip and Arndt 2013; Taillefer and Miller 2014), we take into account only maximal matches no shorter than 30 nucleotide bases. To indicate the result clearly, we exhibit statistics for "forward" and "reverse" strands of intersection/alignment separately, and do statistics in both whole-genome regions and genome regions that are not repeat-masked ("NRM regions"). Percentages in the table represent the conditional probabilities that a maximal match is contained in one of the two sets given that it is contained in the other, weighted by maximal match length. Evidently, *nem* always recovers a large majority of the elements of *net*—P(*nem*|*net*) is always above 90%—but certain of the elements of *nem* are lost to *net* in whole-genome regions. These elements lost to *net* turn out to appear mostly in genome regions that are repeat-masked; in NRM regions, *nem* and *net* are mutually highly consistent with each other. Greater orthology among the compared genomes or genomic regions also increases the consistency between *nem* and *net*; therefore human/chimp shows a higher consistency than human/mouse, and the forward strand of human/chimp shows a higher consistency than the reverse strand.

### 3.3 *Comparison on gene level with orthologous genes annotated by Ensembl Compara*

In this section, we extend inference of lineal orthologs from nucleotide level to gene level. Genes are generally much longer than single maximal matches; a pair of genes may encompass many maximal matches, some non-embedded and others embedded. Following section 2.3 we anticipate that ***each pair of lineal orthologous genes between two genomes shares more non-embedded maximal matches with each other than either of them shares with any other genes in the other genome***, leading naturally to a version of "reciprocal/bidirectional best hit" (RBH) (Tatusov et al. 1997; Bork et al. 1998; Overbeek et al. 1999; Wolf and Koonin 2012). Comparing to single-directional best hits (SBHs), RBHs more faithfully reflect lineal orthologs because SBHs can also be found in collateral orthologs (for example, ***seq2b*** in figure 5 is a SBH for both ***seq1b*** and ***seq1c***, but only the lineal orthologous pair ***seq1b*** and ***seq2b*** constitute a RBH).

Table 2. Comparison between reciprocal best hits of genes (RBH) determined by non-embedded maximal matches and different types of orthologous genes annotated by Ensembl Compara.

| *Genome Pairs* | | Human/Chimpanzee | Human/Mouse |
|---|---|---|---|
| *Total number of RBHs* | | 25159 | 21209 |
| ***Selectivity*** | *P*(*one-to-one* \| *RBH*) | 83.8% (21076/25159) | 67.1% (14225/21209) |
| | *P*(*one-to-many* \| *RBH*) | 2.2% (557/25159) | 4.0% (856/21209) |
| | *P*(*many-to-many* \| *RBH*) | 0.6% (141/25159) | 0.9% (190/21209) |
| | *P*(*otherwise* \| *RBH*)* | 13.5% (3385/25159) | 28.0% (5938/21209) |
| ***Sensitivity*** | *P*(*RBH* \| *one-to-one*) | 94.5% (21076/22303) | 85.0% (14225/16728) |
| | *P*(*RBH* \| *one-to-many*) | 39.1% (557/1424) | 20.7% (856/4130) |
| | *P*(*RBH* \| *many-to-many*) | 16.4% (141/862) | 3.4% (190/5571) |

Note— Probabilities are calculated based on the number of genes or RBHs, without weighting by length in bases. Numbers in brackets exhibit the total numbers of elements contained in the corresponding sets of genes or RBHs.

* This probability denotes the discrepancy between our RBHs and the Ensembl orthologs.

In practice, by using the accumulated amount of non-embedded maximal matches shared between each pair of genes between two genomes as hit score, we seek the RBHs of genes and take them as candidates for lineal orthologs (see section 5.3 for details). For two representative pairs of genomes, human/chimpanzee and human/mouse, we extract such RBHs, and compare them to the orthologous genes obtained for the same genomes from Ensembl Compara. Ensembl classifies three types of orthologs: *one-to-one*, *one-to-many* and *many-to-many* (as shown in figure 1); their relationships with our lineal/collateral orthologs have been discussed in section 2.1. We anticipate most of our RBHs are annotated as orthologs (of any type) in Ensembl, and conversely most of the Ensembl *one-to-one* orthologs are recovered by our RBHs—the former reflects the selectivity while the latter the sensitivity of our method—both as assessed relative to the annotations of Ensembl Compara. Table 2 exhibits the performance of our RHBs recovering Ensembl orthologs: nearly 90% of the RBHs for human/chimpanzee and more than 70% of those for human/mouse are annotated as orthologs in Ensembl, and conversely these RBHs account for nearly 95% of the one-to-one orthologs for human/chimpanzee and 85% of those for human/mouse.

As for one-to-many and many-to-many orthologs, since Ensembl does not provide enough information to distinguish parent and daughter copies of duplications from each other, it is not possible following solely the Ensembl annotations to distill lineal orthologs from all pairs of one-to-many and many-to-many orthologs; the selectivity and sensitivity of our RBHs recovering such orthologs are not readily assessable in this paper.

We also conduct a simple investigation into the discrepancies between our RBHs and the Ensembl orthologs: P(*otherwise*|*RBH*) in table 2 describes the proportion of RBHs that are *not* annotated as orthologs in Ensembl; selectivity of this method can be estimated by its complement. Among all 3385 such RBHs between human and chimpanzee genomes, 2364 of them consist of human and chimpanzee genes that do not appear at all in the Ensembl annotations for gene orthology: orthologs for these genes are not found in Ensembl; thus our inferences of orthology for these genes do not contradict Ensembl. By checking the annotations for these genes, we find in 2140 of these RBHs, either or both genes are annotated as "uncharacterized," or not annotated at all—these RBHs may potentially represent "newly discovered" gene orthology that awaits independent validation; whereas the rest 220 or so mostly (with less than 10 exceptions) consist of human and chimpanzee genes whose annotations are very similar to each other (for example, human gene "*ENSG00000180483*" and chimpanzee gene "*ENSPTRG00000013351*" make a RBH in our calculation but are not annotated as orthologs in Ensembl; however, both these genes are annotated as coding for the "*beta-defensin 119*" protein) and although orthology in principle refers to evolutionary history irrespective of function, RBHs annotated as the same genes in different genomes suggest these genes may be orthologous.

On both nucleotide level and gene level, our method infers *sequence orthology*, which is essentially a relationship between genomic regions, but not necessarily of gene functions. Ensembl annotations of gene orthology incorporate manual curation; some discrepancies between our inferences and Ensembl annotations therefore arise from considerations beyond the scope of automated sequence comparison. For example, different genes within the same genome can overlap, yielding ambiguity when inferring gene orthology via orthology of genomic regions. Supplementary figure 3 shows two examples; in each example, large parts of two human genes overlap with each other, and all non-embedded maximal matches shared with a certain chimpanzee gene appear within the overlapping region of these human genes, suggesting that this overlapping region is orthologous to the chimpanzee gene against which both human genes exhibit the same hit score. Our calculation correctly infers the orthology between genomic regions by recognizing both gene pairs as RBHs; however, based solely on function only one of them is annotated as orthologous in Ensembl. Of the 1021 RBHs of human and chimpanzee genes whose orthologs in Ensembl differ from those inferred by our method, more than 250 arise from sequence overlaps of this form. On the other hand, some Ensembl orthologous genes exhibit poor nucleotide sequence similarity; for example, the human gene "*ENSG00000175505*" and chimpanzee gene "*ENSPTRG00000032521*" are annotated as a pair of one-to-one orthologs in Ensembl, but no matches longer than 20 bases were found between their sequences. This accounts for why some Ensembl one-to-one orthologs are lost in our RBHs; as genomes diverge, the frequencies of such orthologs with low sequence similarity obviously increase.

Updates to Ensembl databases resolve certain discrepancies while introducing new ones. For example, based on genome sequences and gene annotations obtained from version 76 of the Ensembl Core database, our method identified 25129 pairs of human and chimpanzee genes as RBHs; in the corresponding Ensembl Compara database, 3135 (12.5%) of these RBHs were not annotated as orthologs. Of these 3135 pairs of genes, 186 were nevertheless annotated as orthologs in the *succeeding* (current) version 81 of Ensembl Compara. Based on genomic sequences and gene annotations obtained from version 81 of the Ensembl Core database, 184 of these 186 RBHs from version 76 survived.

On the other hand, of the 21994 RBHs annotated as orthologs by Ensembl Compara version 76, 479 were no longer annotated as orthologs in version 81; however, based on version 81 of the Ensembl Core database, 407 of these 479 pairs of human and chimpanzee genes survived as RBHs. Our method could be informative for the calibration of the annotation of gene orthology.

Table 3. Selectivity and sensitivity of a "control experiment," in which RBHs are determined by *all* maximal matches (both embedded and non-embedded).

| *Genome Pairs* | Human/Chimpanzee | Human/Mouse |
|---|---|---|
| *Minimal length of maximal matches*[*] | 60 | 70 |
| *Total number of RBHs* | 24606 | 3813 |
| ***Selectivity:*** *P*(*Ensembl ortholog* \| *RBH*) | 87.7% (21575/24606) | 58.4% (2228/3813) |
| ***Sensitivity:*** *P*(*RBH* \| *one-to-one*) | 93.5% (20863/22303) | 12.5% (2086/16728) |

[*] The performance of such a control experiment depends critically on the minimal length of maximal matches; here we choose the minimal lengths with which these control experiments exhibit the best performances; see supplementary table 1 for control experiments with varying minimal lengths.

For comparison, we also performed a "control experiment:" we extracted RBHs determined by *all* maximal matches (both embedded and non-embedded) and compared its performance with that of our method exhibited in table 2. The performance of such a control experiment depends critically on the minimal length of maximal matches. In table 3, we carefully chose the minimal length that enables the control experiment exhibit its best performance; see supplementary table 1 for control experiments with varying minimal lengths. For human/chimpanzee, the control experiment in table 3 almost exhibits as high performance as our method exhibited in table 2. One possible explanation for this is that between very closely related genomes such as human and chimpanzee, non-embedded maximal matches overwhelmingly dominate embedded ones (see section 3.2.4 in (Gao and Miller 2014)); the *#(nem)/#(em)* values in supplementary table 1 also indicate that when the minimal length is long, an overwhelming majority of hits (matches) between the RBHs of human and chimpanzee consist of non-embedded maximal matches. In this case, RBHs determined by all maximal matches and by solely non-embedded maximal matches are actually nearly equivalent. But for human/mouse, non-embedded maximal matches are not as dominant as for human/chimpanzee. As a result, in the control experiment for human/mouse, we obtained many fewer RBHs; both the selectivity and the sensitivity of these RBHs are greatly reduced. Therefore, the high performance of the calculation in table 2 can't be attributed solely to the RBH approach: *the role of non-embedded maximal matches is essential.*

Although we have discussed here only the human/chimpanzee and human/mouse genomes, we have confirmed that we can also recover with this method orthologous genes from other mammalian genome pairs with similarly high selectivity and sensitivity. For more distantly related genome pairs, e.g. mammals versus birds or fish, the increase of orthologous genes with low sequence similarity (as we discussed above) causes methods based solely on comparison of nucleotide sequences to gradually lose their effectiveness; however, RBHs determined by non-embedded maximal matches still show remarkably higher selectivity and sensitivity in recovering orthologs than those determined by all maximal matches. This evidence supports our proposal in section 2.3 about the relationship between non-embedded maximal matches and lineal orthologs; data will be reported elsewhere.

# 4. Discussion

## 4.1 *Concepts of one-to-one orthology*

The term "orthologs" is often used to refer to "equivalent genes" (Kuzniar et al. 2008, Dewey 2011). Nevertheless, the existence of co-orthologs does suggest that "orthologs" alone does not necessarily recapitulate the meaning of the phrase "equivalent genes:" in-paralogs are not always equivalent to their common orthologous cognates (Dewey 2011); co-orthologs must be further classified to isolate those "most orthologous" pairs of genes that can be expected to faithfully represent the notion of "equivalent genes."

By design, graph-based methods of ortholog identification yield one-to-one (or nearly one-to-one) orthologs based on certain *working* definitions of sequence orthology; the one-to-one orthologs so obtained constitute a proper subset of all orthologs by customary definitions based on phylogeny. However, the working definitions alone do not faithfully represent the evolutionary significance of these one-to-one orthologs; *conceptual* definitions are therefore required for validation of these orthologs by phylogeny.

The first conceptual definition to capture the gist of the problem might have been "positional ortholog" as articulated by (Koski et al. 2001) and a little bit later by (Swidan et al. 2006); a similar notion of "positional homolog" was proposed by (Bourque et al. 2005) and (Burgetz et al. 2006). Swidan *et al.* remark that "positional orthologs are orthologs that have preserved their relative positioning or genomic contexts in the genomes," suggesting a criterion to assess positional orthology, namely context—the sequences of their flanking regions. Subsequently, Dewey offered a more systematic and precise definition of positional orthology (Dewey 2011). Dewey's definition elucidates the relationship between positional orthology and parent/daughter copies of duplications; the distinction between parent and daughter copies of duplications is essential to positional orthology. Dewey also accounted for what he called the "symmetry" of duplications. Positional orthology is defined only for asymmetrical duplications, because for symmetrical duplications the parent and daughter copies are indistinguishable; therefore, positional orthology is not strictly one-to-one (Dewey 2011).

Later, Han and Hahn defined a "primary ortholog" as the ortholog between parent copies of duplications (Han and Hahn 2009); however, the authors didn't account for the distinction between ancient duplications and recent ones. From our perspective, orthology between the daughter copies of an ancient duplication ought to be primary (see figure 1 (a) and figure 3 (a))—but this is not articulated in (Han and Hahn 2009). We therefore eschew the term "primary ortholog" in favor of "lineal ortholog" instead.

In this paper, we propose the novel concept of "lineal orthology" based on the "direct ancestor" of each genomic sequence; this relation is transitive and strictly one-to-one. With our definition, the parent and daughter copies of a duplication can be readily distinguished: the parent copy has a longer history of "direct inheritance" than the daughter copy, whereas the history of the daughter copy can't be traced to any time before the duplication. In contrast to Dewey's definition of "positional orthology," our definition does not account for the symmetry of duplication; in principle, parent and daughter copies of a symmetrical duplication also have distinct histories of "direct inheritance"—although an efficient technique of determining these histories is still to be developed. For the calculations reported in this paper, the impact of symmetrical duplication is minor.

In (Gao and Miller 2014), we reported the observation of an exponential length distribution of all exact matches between human and chimpanzee genomes; these exact matches can be split into two subsets. One subset retains the exponential length distribution of the whole set, while the other subset exhibits a power-law length distribution with exponent -3. In (Gao and Miller 2014), we suggested that the former subset consists primarily of orthologs and the latter of paralogs, whereas in the viewpoint of this paper, it is *lineal* orthologs that primarily compose the exponential length distribution, and collateral orthologs plus paralogs that chiefly comprise the power-law length distribution. These observations are consistent with—although far more specific than—those of Arndt and co-workers (Massip et al. 2015).

## 4.2 *Synteny, orthology and genomic context*

The term "synteny" is widely used to refer to homologous genomic regions with conserved gene order. Although this colloquial application of the term "synteny" is not consistent with its original definition (Renwick 1971; Passarge et al. 1999), "synteny" is commonly taken as synonymous with "orthology," leading for example to the failure to distinguish a "synteny map" from an "orthology map." The promiscuous usage of this term could reflect a potential relationship in practice between synteny—in the sense of "conserved gene (or sequence) order"—and orthology. It has been observed in previous studies that synteny-based inference of orthology yields high concordance with sequence-based inference of orthology, illustrating that "local synteny is a robust substitute to coding sequence for identifying orthologs" (Jun et al. 2009a).

Synteny information has been widely used for predicting or refining orthologous relationships. Synteny can refer to genomic context over relatively long regions that may span a series of genes; in previous literature, "genomic context" generally means "synteny." Dewey recognized the biological significance of genomic context, and classified "orthology prediction methods that take genomic context into account" into three categories: (i) methods based primarily on sequence similarity but that also incorporate conserved gene order or conserved gene neighborhoods; (ii) methods that combine sequence similarity with gene order evolutionary models, and a parsimony principle; and (iii) methods incorporating synteny information, including synteny block generators and whole-genome alignment tools (Fu et al. 2007; Dewey 2011). Obviously, these methods (i), (ii) and (iii) are all based on synteny information—information on genomic context over relatively long ranges. Our work in this paper for the first time extends the idea of context-based ortholog identification from long-ranged contexts to short-ranged ones—as short as single maximal matches—that we demonstrate provide an effective means to identify orthologs within whole genomes. The method of non-embedded maximal matches proposed in this paper relies on local embedding of sequences and synteny information plays no direct role; its overall agreement on the nucleotide level with a Lastz net alignment (see section 3.2) therefore suggests a relationship between synteny and homology: although our non-embedded maximal matches are not defined in terms of synteny, a large majority of the in-synteny sequences turn out to be non-embedded.

## 4.3 *Intersection and alignment*

Genome alignment arranges sequences to identify regions of similarity that may have arisen from evolutionary relationships (Mount 2004). A typical alignment consists of two phases: an alignment-search phase that includes an all-against-all search for identical or similar sequences, and a subsequent mapping/clustering phase based on these sequences. Intersection is one option for the first phase of alignment; a basic intersection exhaustively recovers all sequences common to the compared genomes, without any filtering or rearrangement. Alignment methods that perform intersection as a first step include, for example, *MUMmer* (http://mummer.sourceforge.net/). The term "intersection" was applied in (Salerno et al. 2006); elsewhere, e.g. in (Massip and Arndt 2013), it is called "alignment" although according to our use of the term it is only a preliminary stage thereof and would require in addition some form of mapping or clustering to qualify as an alignment.

For many purposes including our own the distinction between intersection and alignment is important. Our method based on non-embedded maximal matches, although it can be performed on an alignment (see section 3.2.4 in (Gao and Miller 2014)), doesn't require alignment to identify and classify orthologs; intersection is sufficient. In contrast to alignment-based methods that are algorithmically defined, intersection-based methods like ours exhibit certain virtues. Intersection involves few parameters; the objects computed can be easily and precisely described. This reduces the possibility of artifacts due to the choice of parameters and details of alignment algorithm. With existing computational technology, intersection is easier to perform than many widely used alignment methods such as Blastz/Lastz; for a given sequence pair, intersection usually consumes less computation time than alignment, particularly for whole-genome comparison. For example, with a single core of Intel Xeon E5-2680v3 processor at 2.50 GHz on our high performance computing cluster, an intersection with *SEQANALYSIS* together with full classification of super, nested local and non-nested local maxmers (see section 5.2 and (Taillefer and Miller 2014) for details) between human chromosome 1 and chimpanzee chromosome 1, neither of which were repeat-masked, requires around 20 minutes of computation time; in contrast, a raw alignment with Lastz for the same pair of chromosomes—but repeat-masked—takes about 4 hours. When comparing whole genomes, an intersection between 3.1 GB of human sequence and 3.2 GB of chimpanzee sequence with *SEQANALYSIS* requires less than 24 hours, whereas the corresponding Lastz raw alignment can take more than 720 hours of computation time with a single core of CPU on our cluster. This also leads to a result that intersection can be implemented on a wider range of sequences. Blastz/Lastz alignment requires prior repeat-masking of the sequences to be aligned; however, intersection can be implemented directly on whole-genome sequences as we have done for computations described in this paper, without any repeat-masking.

## 4.4 *Conclusion and Outlook*

The customary definition of "co-orthologs" notwithstanding, there is a certain consensus on the practical utility of a potential "one ortholog one organism" relationship (Han and Hahn 2009; Dewey 2011). Among a group of co-orthologous genes, we assume that there is one and only one "most orthologous" gene pair that retains its common ancestor's position in the genome, and that the genes comprising this pair are more likely to play equivalent roles within both genomes than are other homologs; it is an empirical matter whether this assumption is borne out in practice. We introduced here the notion of "lineal ortholog" to describe this "most orthologous" relationship among genes or sequences. Lineal orthologs are defined as orthologs that have the same direct ancestor; orthologs having distinct direct ancestors are designated as "collateral orthologs." The evolution of lineal orthologs is presumably constrained primarily by negative selection, whereas the evolution of collateral orthologs evidently more resembles that of paralogs. These observations are consistent with those of Arndt and coworkers (Massip et al. 2015).

Lineal orthologs can be identified by information in their flanking regions—as we have proposed in this paper, the local structure of embedded/non-embedded maximal matches can efficiently discriminate lineal orthologs from other homologs. The calculation can be done with either a genomic intersection or a genomic alignment; the intersection-based calculation involves fewer parameters—so that the computed objects are simple to elucidate—and is less computationally intensive than commonly used alignment-based methods. Combined with a reciprocal best hit approach, non-embedded maximal matches elucidate lineal orthologs not only for sequences but also for genes. Reciprocal best hits of genes containing non-embedded maximal matches recover orthologous genes with both high selectivity and high sensitivity; the inferred orthologous genes are quite consistent with the annotation of one-to-one orthologs in Ensembl Compara. Moreover, non-embedded maximal matches always recover from the compared genomes a large majority of the sequences extracted by Lastz net alignment; this consistency suggests a potential relationship between orthology and synteny. Inasmuch as relatively short contigs alone should suffice for the intersection-based computations reported here, a prospect to be pursued elsewhere is raised that for purposes of identifying and inferring evolutionary history of orthologs genome-wide, it may eventually be possible to bypass or significantly abridge the process of genome assembly.

# 5. Materials and Methods

## 5.1 *Genomic data downloaded from Ensembl*

For our intersections we download whole-genome sequences of human (ftp://ftp.ensembl.org/pub/release-81/fasta/homo_sapiens/dna/Homo_sapiens.GRCh38.dna.toplevel.fa.gz), chimpanzee (ftp://ftp.ensembl.org/pub/release-81/fasta/pan_troglodytes/dna/Pan_troglodytes.CHIMP2.1.4.dna.toplevel.fa.gz) and mouse (ftp://ftp.ensembl.org/pub/release-81/fasta/mus_musculus/dna/Mus_musculus.GRCm38.dna.toplevel.fa.gz) from the ftp server of Ensembl (http://www.ensembl.org/index.html). In this paper, we use version 81 (the latest version available in September 2015) of the Ensembl databases.

Gene annotations for all species can be obtained from the Ensembl core database. We extract the information we need from the corresponding tables of the MySQL server at Ensembl (ftp://ftp.ensembl.org/pub/release-81/mysql/). Annotations of gene orthology can be obtained from the Ensembl Compara database; tables for the Compara database version 81 can be downloaded from ftp://ftp.ensembl.org/pub/release-81/mysql/ensembl_compara_81/. We wrote simple code to read these tables and extract the entire list of Ensembl annotated orthologous genes. Alternatively, one can use the Ensembl APIs (http://www.ensembl.org/info/docs/api/index.html) to fetch the same data programmatically.

For the whole-genome pairwise alignments in section 3.2, we downloaded the compressed MAF files from the Ensembl ftp server: ftp://ftp.ensembl.org/pub/release-81/maf/ensembl-compara/pairwise_alignments/homo_sapiens.GRCh38.vs.pan_troglodytes.CHIMP2.1.4.tar for the whole-genome alignment of human/chimpanzee, and ftp://ftp.ensembl.org/pub/release-81/maf/ensembl-compara/pairwise_alignments/homo_sapiens.GRCh38.vs.mus_musculus.GRCm38.tar for that of human/mouse. Methods of extracting all exact matches from these alignments are described in (Gao and Miller 2011) and (Gao and Miller 2014).

## 5.2 *Embedded and non-embedded maximal matches in a whole-genome pairwise intersection*

To obtain all embedded and non-embedded maximal matches between two genomes, we first perform an "intersection" to collect all sequences shared by these genomes (Salerno et al. 2006). Of these shared sequences, maximal matches are most often of primary interest, and in this paper "intersection" refers to the *complete set* of maximal matches above some specified length in bases. Embedded and non-embedded maximal matches can be directly classified from the complete list of maximal matches according to their definitions in section 2.2. However, without special techniques, both the intersection and the classification of embedded/non-embedded maximal matches are formidable tasks for large datasets.

Advances in hardware technology and data structures have made it increasingly practicable to perform a whole-genome intersection. A widening variety of algorithms and software has emerged, many based on a *suffix tree* or *suffix array*, for example, *MUMmer* (http://mummer.sourceforge.net/) (Delcher et al. 1999; Delcher et al. 2002; Kurtz et al. 2004) and *SEQANALYSIS* (Taillefer and Miller 2011a; Taillefer and Miller 2014). Such data structures can be built and searched in linear time and linear space. Without loss of generality, we use *SEQANALYSIS* in this paper to calculate our intersections. *SEQANALYSIS* organizes its output in a compact way: instead of a complete list of maximal matches, it generates a complete list of "*maxmers*." Each maxmer consists of a set of *occurrences*, sequences with identical content in each of the compared genomes. Each occurrence in one genome contributes to a match—but not necessarily a "maximal" match—with each occurrence in the other genome; for each maxmer, there is at least one pair of occurrences—one from each genome—that constitutes a maximal match (see details in (Taillefer and Miller 2014)).

Table 4. Correspondence among occurrences of maxmers and non-embedded/embedded maximal matches.

<table>
<tr><td rowspan="2">Occurrences of maxmers</td><td>super maxmer</td><td>(a) unique super or (b) non-unique super</td></tr>
<tr><td>local maxmer</td><td>(c) non-nested local or (d) nested local</td></tr>
<tr><td rowspan="2">Non-embedded maximal matches (nem)</td><td>MUMs</td><td>(1) two unique supers</td></tr>
<tr><td>nem but not MUMs</td><td>(2) a unique super + a non-unique super, or vice versa<br>(3) two non-unique supers<br>(4) two non-nested locals</td></tr>
<tr><td colspan="2">Embedded maximal matches (em)</td><td>(5) a non-nested local + a nested local, or vice versa<br>(6) two nested locals, if they make a maximal match</td></tr>
</table>

Note— Maxmers are classified as "super" or "local;" occurrences of super maxmers are subclassified as "*unique super*," if they have exactly one occurrence in the corresponding genomes, or "*non-unique super*," if they have multiple numbers of occurrences in the genomes; occurrences of local mamxers are subclassified as "*non-nested*" or "*nested*" as in (Taillefer and Miller 2014). Among them, *super* and *local* are exclusive of each other, therefore these four types of maxmer occurrences make only six different combinations; four of them are non-embedded maximal matches and two of them embedded maximal matches.

*SEQANALYSIS* also provides an expedient way to quickly identify embedded and non-embedded maximal matches, classifying all maxmers into "*super*" or "*local*," in which the occurrences of the latter can be subclassified into "*nested*" and "*non-nested*;" for details see (Taillefer and Miller 2011b; Taillefer and Miller 2014). Very recently, an algorithm was developed to perform the context-sensitive maxmer classification in linear time (Ohlebusch and Beller 2014). "Non-nested and nested *occurrences*" of local maxmers defined in (Taillefer and Miller 2014) differ from "non-embedded and embedded *maximal matches*" as we define them here: an *occurrence* is a *single* sequence, whereas a *maximal match* consists of a *pair* of sequences. Our non-embedded maximal matches consist of super maxmers or non-nested occurrences of local maxmers; on the other hand, a maximal match is embedded if at least one of the two sequences that constitute it is a nested occurrence of local maxmer; see table 4 for their correspondences. This allows us to identify non-embedded and embedded maximal matches independently (see supplementary material 1 for details) and greatly reduces the computational burden.

In this paper, we retrieve with *SEQANALYSIS* all occurrences of "4-base maxmers" from the compared genomes, which are not repeat-masked. For the numerical simulations in section 3.1, we only make the intersections between the forward strands of synthetic sequences; while in section 3.2 and 3.3, we make intersections for both the forward and the reverse complement strands of the compared genomes. Non-embedded and embedded maximal matches are obtained from these maxmer occurrences according to table 4; see supplementary material 1 for computation details.

Where indicated, we also examine another set of maximal matches—maximal unique matches ("*MUMs*" for short); *MUMs* form a proper subset of all non-embedded maximal matches defined in this paper (see table 4). *MUMs* can be identified either by *SEQANALYSIS* according to table 4, or by *MUMmer* (http://mummer.sourceforge.net/) with switch *–mum*; the latter also provides another switch *-maxmatch*, which returns a complete list of all maximal matches, irrespective of embedding. For whole-genome scale computations, the reduction of the list of all maximal matches returned by *MUMmer* to the lists of non-embedded (or embedded) maximal matches returned directly by *SEQANALYSIS*, is not feasible on realistic time scales.

## 5.3 *Reciprocal best hits (RBHs) of genes determined by a given group of maximal matches*

Given the region of a gene within a whole genome, it is straightforward to identify maximal matches that overlap with that gene region. We retrieve "reciprocal best hits" (RBHs) of genes from the two genomes based on their overlap of each gene with a given group of maximal matches. Overlap between a single maximal match and a pair of genes is illustrated in figure 7: ***gene1*** occupies a region [***start1***, ***end1***] in ***Genome1***, and ***gene2*** occupies a region [***start2***, ***end2***] in ***Genome2***. A maximal match consisting of a subsequence [***x1***, ***x2***] of ***Genome1*** and a subsequence [***y1***, ***y2***] of ***Genome2***, overlaps with these two genes. We define a pair of subsequences with equal lengths, [*max*{***x1***, ***start1***, ***start2***+***x1***-***y1***}, *min*{***x2***, ***end1***, ***end2***+***x2***-***y2***}] in ***Genome1*** and [*max*{***y1***, ***start2***, ***start1***+***y1***-***x1***}, *min*{***y2***, ***end2***, ***end1***+***y2***-***x2***}] in ***Genome2*** as the "*matched region*" between ***gene1*** and ***gene2*** associated to this single maximal match (red bars in figure 7). The "hit" score for a pair of genes is the sum, over a selected subset of maximal matches between those genes, of the lengths of all matched regions associated to a given maximal match. For each gene in one genome, the gene in the other genome that shares the greatest hit score with it is defined as its (*single-directional*) *best hit*. If two genes are best hits of each other, then this pair of genes is defined as a *reciprocal/bidirectional best hit* (RBH) of genes, determined by the given group of maximal matches.

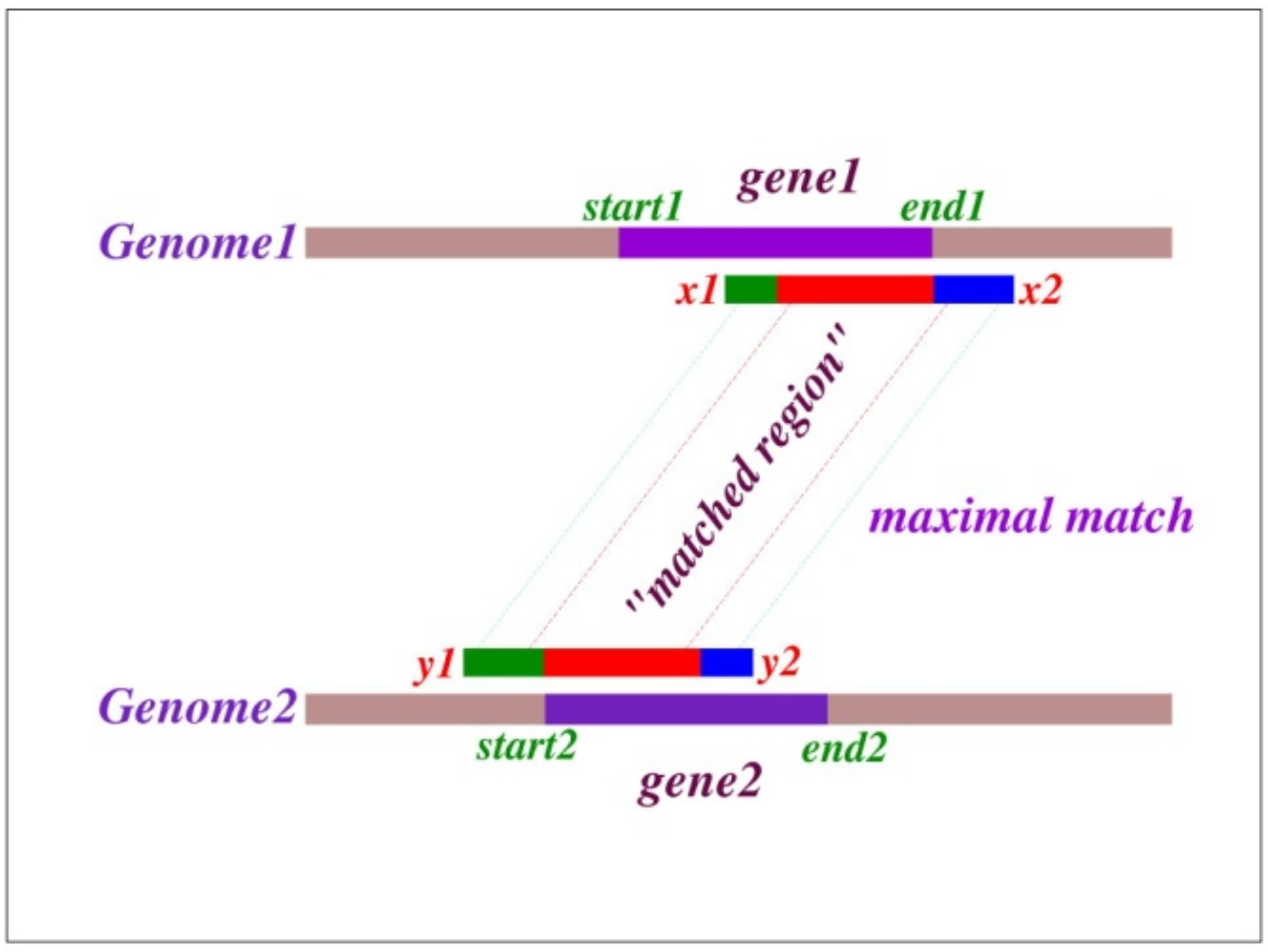

Figure 7. Matched region between a pair of genes associated to a given maximal match.

The selected subset of maximal matches determines the RBHs. In section 3.3, we take the RBHs determined by the *non-embedded* maximal matches between two vertebrate genomes as candidates for lineal orthologous genes, and the RBHs determined by *all* maximal matches between these two genomes as a null reference. Locations of genes in the genomes are obtained from the Ensembl annotations for gene regions; maximal matches (non-embedded or embedded) between whole-genome sequences are obtained through whole-genome intersections (see section 5.2 and supplementary material 1).

## Data Access

Source code of our software package, *SEQANALYSIS*, can be obtained in the webpage of Physics and Biology Unit, Okinawa Institute of Science and Technology Graduate University (https://groups.oist.jp/sites/default/files/imce/u109/sequanalysis.zip). A full instruction manual is also available with the package. All data sources for the calculations in this paper are publicly available.

## Acknowledgement

This work was supported by the Physics and Biology Unit of Okinawa Institute of Science and Technology Graduate University. The authors gratefully acknowledge Dr. Oleg Simakov, Dr. Eddy Taillefer, Dr. Zdenek Lajbner and Mr. Quoc Viet Ha for their valuable input, stimulating discussions and technical support on this paper.

## Supplementary Figures

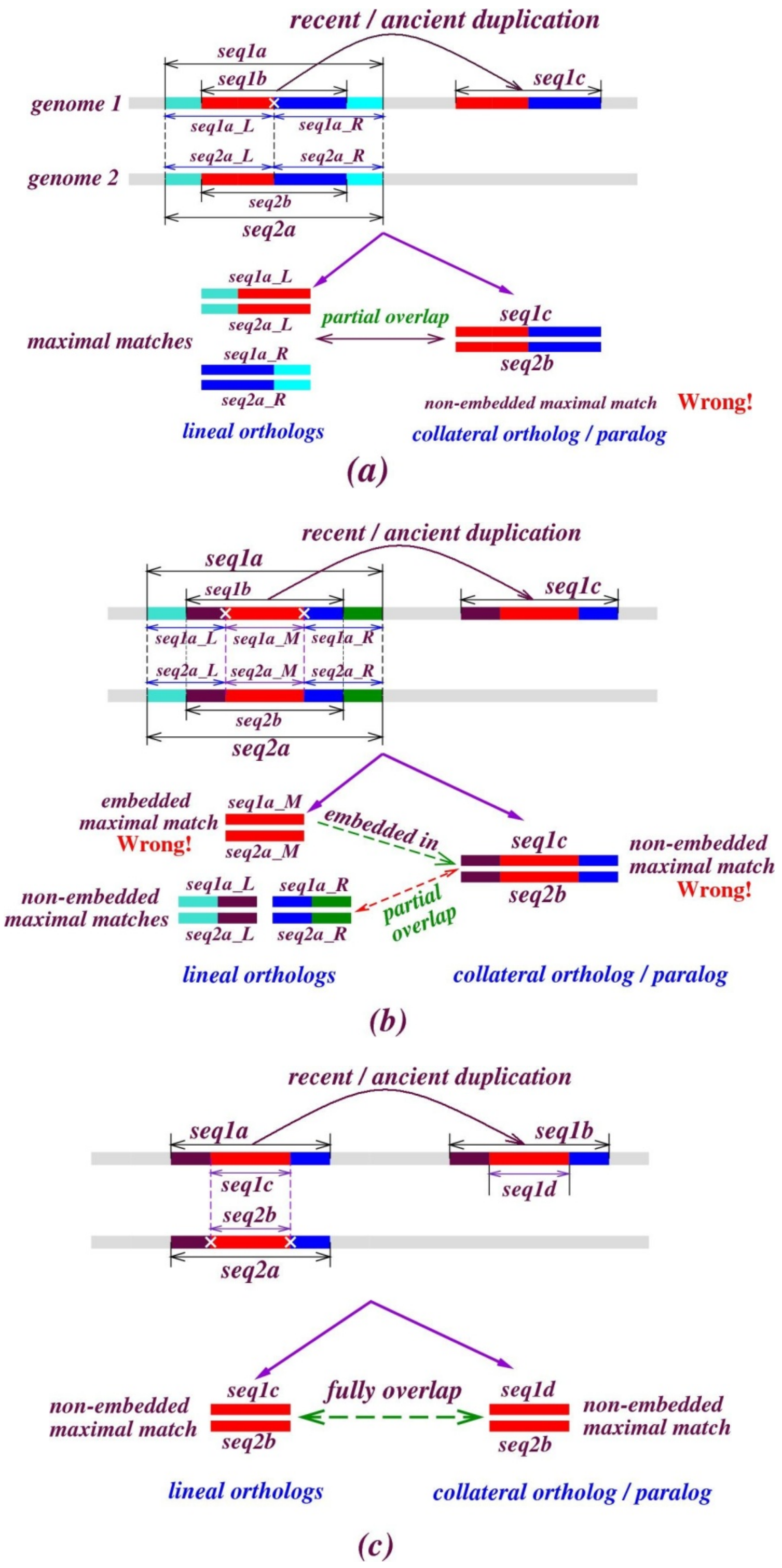

Supplementary figure 1. Substitutions (shown in white "X"s) in the parent copy of segmental duplication may yield misclassification of lineal orthologs from collateral orthologs and paralogs.

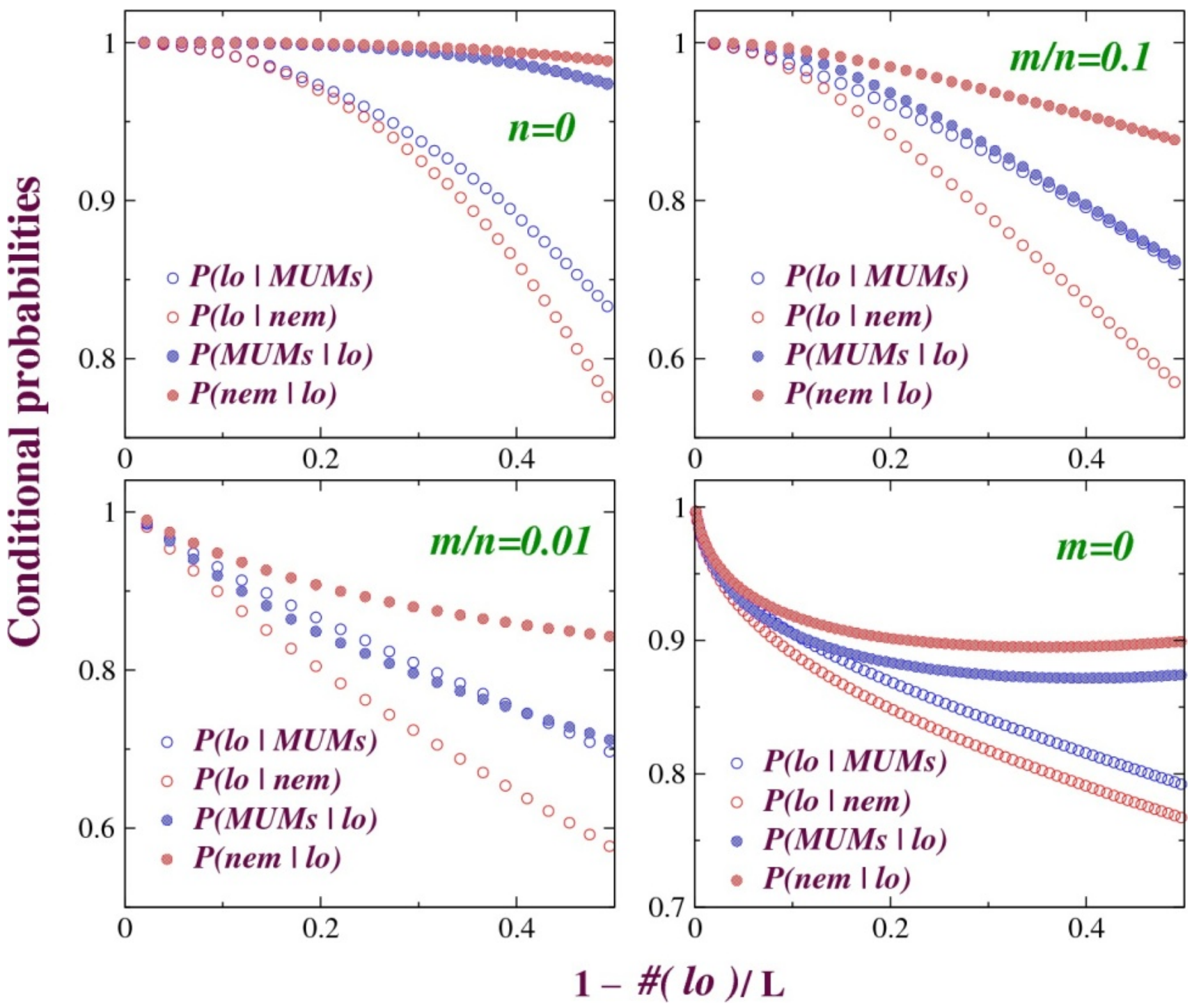

Supplementary figure 2. Conditional probabilities showing the selectivities and sensitivities of non-embedded maximal matches (*nem*) and maximal unique matches (*MUMs*) recalling lineal orthologs (*lo*) in the numerical simulations in section 3.1. Comparing to *nem*, *MUMs* have higher selectivities (P(*lo*|*MUMs*)>P(*lo*|*nem*)) but lower sensitivities (P(*MUMs*|*lo*)<P(*nem*|*lo*)).

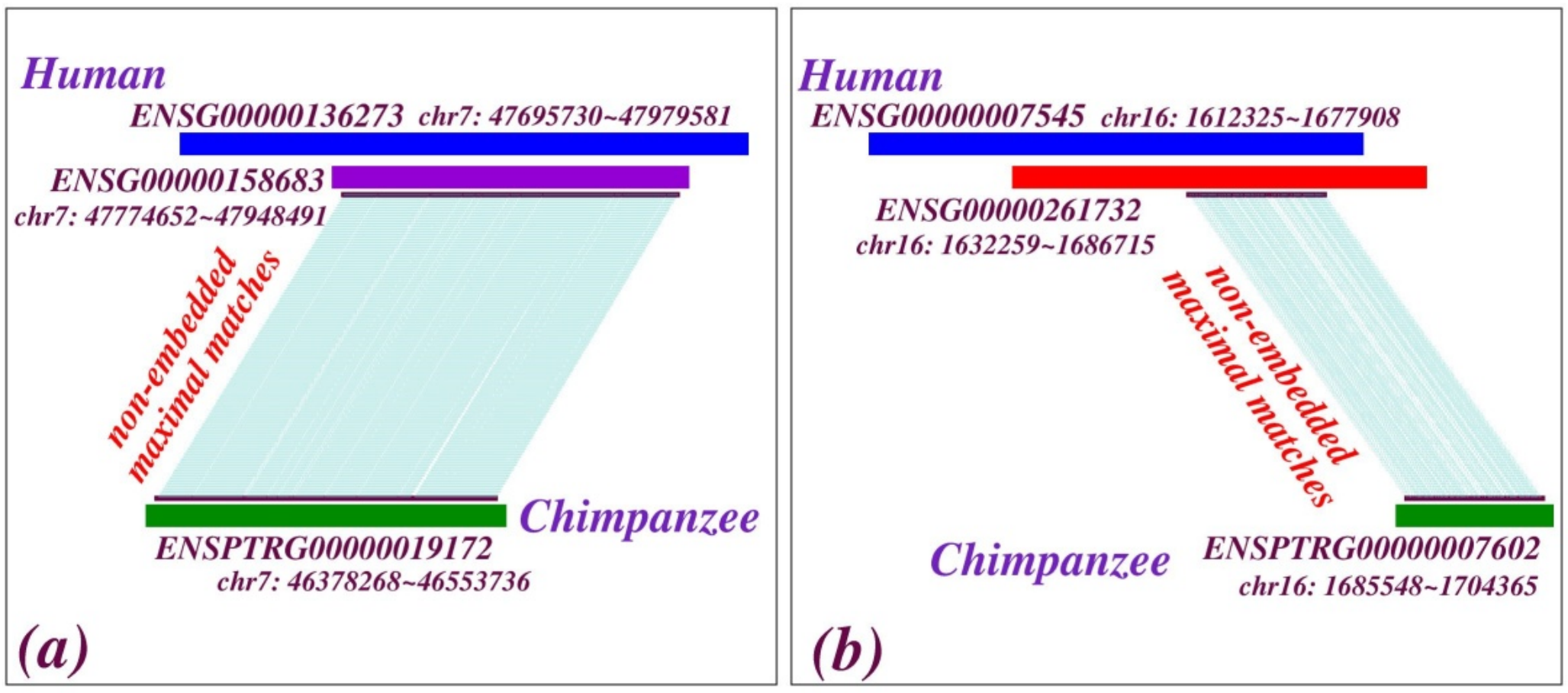

Supplementary figure 3. Major overlaps between genomic regions of different genes may complicate inference of gene orthology.

# Supplementary Table

| Minimal length | Maximal matches | Total number of RBHs | #(nem)/#(em) | ***Selectivity*** *P*(*Ensembl ortholog* \| *RBH*) | ***Sensitivity*** *P*(*RBH* \| *one-to-one*) |
|---|---|---|---|---|---|
| 20 | *nem+em* | 8023 | 0.91 | 79.4% (6372/8023) | 26.9% (6005/22303) |
| | *nem* | 25159 | | 86.5% (21774/25159) | 94.5% (21076/22303) |
| 30 | *nem+em* | 16593 | 6.99 | 87.4% (14500/16593) | 62.7% (13983/22303) |
| | *nem* | 25173 | | 86.6% (21794/25173) | 94.6% (21094/22303) |
| 40 | *nem+em* | 23480 | 17.8 | 88.6% (20806/23480) | 90.3% (20145/22303) |
| | *nem* | 25181 | | 86.6% (21809/25181) | 94.6% (21102/22303) |
| 50 | *nem+em* | 24321 | 32.9 | 88.0% (21397/24321) | 92.8% (20708/22303) |
| | *nem* | 25127 | | 86.7% (21775/25127) | 94.5% (21078/22303) |
| 60 | *nem+em* | 24606 | 47.3 | 87.7% (21575/24606) | 93.5% (20863/22303) |
| | *nem* | 25057 | | 86.8% (21751/25057) | 94.4% (21045/22303) |
| 70 | *nem+em* | 24617 | 61.7 | 87.4% (21520/24617) | 93.2% (20797/22303) |
| | *nem* | 24897 | | 86.9% (21628/24897) | 93.8% (20918/22303) |
| 80 | *nem+em* | 24507 | 76.4 | 87.3% (21402/24507) | 92.7% (20683/22303) |
| | *nem* | 24708 | | 86.9% (21471/24708) | 93.1% (20769/22303) |

(a) For human/chimpanzee

| Minimal length | Maximal matches | Total number of RBHs | #(nem)/#(em) | ***Selectivity*** *P*(*Ensembl ortholog* \| *RBH*) | ***Sensitivity*** *P*(*RBH* \| *one-to-one*) |
|---|---|---|---|---|---|
| 20 | *nem+em* | 3134 | 0.03 | 38.2% (1196/3134) | 5.6% (933/16728) |
| | *nem* | 21209 | | 72.0% (15271/21209) | 85.0% (14225/16728) |
| 30 | *nem+em* | 4158 | 0.11 | 42.9% (1785/4158) | 9.0% (1509/16728) |
| | *nem* | 17957 | | 67.9% (12197/17957) | 68.0% (11371/16728) |
| 40 | *nem+em* | 4180 | 0.57 | 45.7% (1911/4180) | 10.1% (1694/16728) |
| | *nem* | 13615 | | 64.6% (8797/13615) | 49.4% (8267/16728) |
| 50 | *nem+em* | 3939 | 0.89 | 52.2% (2058/3939) | 11.4% (1899/16728) |
| | *nem* | 10338 | | 61.8% (6384/10338) | 36.3% (6067/16728) |
| 60 | *nem+em* | 3863 | 1.71 | 57.3% (2213/3863) | 12.4% (2073/16728) |
| | *nem* | 7992 | | 58.9% (4704/7992) | 26.8% (4491/16728) |
| 70 | *nem+em* | 3813 | 2.16 | 58.4% (2228/3813) | 12.5% (2086/16728) |
| | *nem* | 6216 | | 58.9% (3660/6216) | 20.9% (3489/16728) |
| 80 | *nem+em* | 4026 | 2.74 | 51.8% (2085/4026) | 11.8% (1973/16728) |
| | *nem* | 4995 | | 57.8% (2886/4995) | 16.5% (2765/16728) |

(b) For human/mouse

Supplementary table 1. Selectivity and sensitivity of RBHs with respect to Ensembl Compara, determined by (i) both non-embedded and embedded maximal matches ("*nem+em*") and (ii) non-embedded maximal matches only ("*nem*"). *#(nem)/#(em)* indicates the rate of hits (matches) contributed by non-embedded maximal matches over those contributed by embedded maximal matches. For human/chimpanzee, when the minimal length of maximal matches is long, non-embedded maximal matches overwhelmingly dominate embedded ones; RBHs determined by (i) and (ii) are nearly equivalent. For human/mouse, non-embedded maximal matches are not as dominant as for human/chimpanzee; with increasing minimal length, while *#(nem)/#(em)* also increases somewhat, more orthologous elements get lost. Therefore, for human/mouse, RBHs determined by (ii) with short minimal length performs the best; RBHs determined by (i) exhibit much lower selectivity and sensitivity than those determined by (ii).

# Supplementary materials

## ***Supplementary material 1:*** *How to identify non-embedded and embedded maximal matches with SEQANALYSIS*

*SEQANALYSIS* is a suffix array-based software package for quickly identifying certain classes of context-sensitive maxmers from pairs of eukaryote genome-length character sequences (Taillefer and Miller 2011a; Taillefer and Miller 2011b). Source code can be obtained from (https://groups.oist.jp/sites/default/files/imce/u109/sequanalysis.zip).

To identify non-embedded and embedded maximal matches between two genomic sequences, the first step is to generate an intersection. *SEQANALYSIS* provides options that return in the output all *super*, *nested local* and *non-nested local* maxmers (see (Taillefer and Miller 2014) for details). We use the following command line for the intersections in this paper:

```
user@server:DIR> seqanalysis --mode=2SeqCountMtch --min-length=<minimal length of maxmers>
                 -L -P -p -fp -A -U -D -n -K -ao --ref-file=<fasta file for the reference sequence>
                 --query-file=<fasta file for the query sequence> --output-file=<output file>
```

where the options represent:

| | |
|---|---|
| --mode=2SeqCountMtch | Pairwise intersection computation. |
| --list-match [-L] | List positions of all occurrences. |
| --print-matchseq [-P] | Print match strings. |
| --print-compact [-p] | List positions of all occurrences compactly. |
| --forward-pos [-fp] | For occurrence on the reverse strand, record the coordinate by its start position on the reverse strand, lead by a "-". For example, -3 means the start position of an occurrence is at the third base of the reverse strand. |
| --max-match [-A] | Compute all maximal matches, irrespective of whether or not they are unique in either genome. |
| --uppercase [-U] | Pre-process the sequences by converting all alphabetical letters to uppercase. |
| --type-dna [-D] | Pre-process the sequences by converting all symbols other than {A, a, T, t, G, g, C, c, N, n} to "N" or "n". |
| --cal-overlap [-n] | Output only super maxmers and non-nested occurrences of local maxmers. |
| --min-length= | Minimal match length. |
| --ref-file= | Input the reference sequence in fasta format. When the input file contains multiple sequences, *SEQANALYSIS* will preprocess these sequences by concatenating them all into a single long sequence, putting a "$" between neighboring sequences. Positions of maxmer occurrences in the output are reported relative to the single long sequence. |
| --query-file= | The same as *--ref-file* but for the query sequence. |
| --output-file= | Name of the output file. |

Optional switches:

| | |
|---|---|
| --append-revcomp [-K] | Compute intersections for both the forward strand and the reverse strand. |
| --print-allocc [-ao] | Print all maxmer occurrences (not only *super* and *non-nested local* but also *nested local*); -ao must be used together with –n. |

For more details, refer to *Manual.pdf* included in the source code of *SEQANALYSIS*.

The output file includes a list of all maxmer occurrences; non-embedded and embedded maximal matches can be obtained from the list of maxmer occurrences according to table 4 in section 5.2 of the main text.

## *S1.1 When only non-embedded maximal matches are needed*

When only non-embedded maximal matches are needed, we can turn off the switch *–ao* to ignore all occurrences of *nested local* in the intersection; this simplifies both the intersection and the identification of non-embedded maximal matches. The output file reads like:

```
1  S:40608 114:AAAAAAAAAAAAAAAAAAAAAAAAAAAAAAAAAAAAAAAATTAATACATGATTTGATTTTAAATATTAAATGACTTTCTTTTT
   ATTTTCCTTCTCTTCTCTCTCACATTCTTAA 1:808567429 2:786292654
2  S:13896708 110:AAAAAAAAAAAATTAGCCGGGCATGGTGGCGGGCGCCTGTAGTCCCAGCTACTCGGGAGGCTGAGGCAGGAGAATGGCGT
   GAACCCGGGAGGCGGAGCTTGCAGTGAGCC 1:1069760770 1:-2726413043 2:205357112
3  L:38942624 101:AAAAAAATACGAAAACCAGTCAGGCGTGGCGGCGTGCGCCTGCAATCGCAGGCACTCGGCAGGCTGAGGCAGGAGAATCA
   GGCAGGGAGGTTGCAGTGAGC 1:1828466975 2:1340474564
4  L:806971 102:AAAAAAAAAAAAAAAAAAAAAAAAAGACTGTAGGAGCATCTGGTGGGAGGTGGTGGAGGGAGAACTGTGGGTTTGGAAGCTG
   CGCCCTCCCCCCAGCCATGC 2:2597025985
5  L:1004765 106:AAAAAAAAAAAAAAAAAAAAAAAAGAAGGAAGGAAGGGCCCAGAAGTCAGGAAGGAGCACGTGAGGAGGGTGTGTGGGAAG
   AATGGAGGTACTGAGGCAGGGTGCA 1:-2504991767
6  L:689564 110:AAAAAAAAAAAAAAAAAAAAAAAAAAGTCAGGAAACAACAGGTGCTGGAGAGGATGTGGAGAAATAGGAACACTTTTACACT
   GTTGGTGGGACTGTAAACTAGTTCAACC 1:-2897341951 1:-2290382390 1:-2216188336 1:-1822542408 2:3166049008
7  L:14377644758 172:GTTGTTGAATTTTGTCAAAGGCCTTTTCTGCATCTATTGAGATAATCATGTGGTTTTTGTCTTTGGTTCTGTTTATA
   TGCTGGATTACGTTTATTGATTTTCATATGTTGAACCAGCCTTGCATCCCAGGGATGAAGCCCACTTGATCATGGTGGATAAGCTTTTTGATGTG
   1:1135681928 2:-2709643878
```

Each line exhibits a maxmer. The left-most column shows the line numbers. The first column to the right indicates the maxmer type ("S" for super maxmer and "L" for local maxmer) and id # (a unique integer for each maxmer), separated by a colon; the second column indicates the length and match sequence, also separated by a colon; and the remaining columns indicate positions of all occurrences of *super* or *non-nested local* in both reference and query sequence ("1:" for occurrence in the reference sequence, and "2:" for occurrence in the query sequence; integers prefixed by "-" indicate that the corresponding occurrences appear in the reverse strand); columns are separated by spaces.

According to table 4 in section 5.2 of the main text, for each maxmer appearing in the output file, since all these occurrences are either *super* or *non-nested local*, each occurrence in the reference sequence and each occurrence in the query sequence constitutes a non-embedded maximal match; due to the same reason, every pair of occurrences must form a maximal match—there is no need to confirm this by checking their contexts.

In the sample output above, all lines except line 4 and 5 indicate non-embedded maximal matches: line 4 contains no (non-nested) occurrence in the reference sequence, while the line 5 contains no (non-nested) occurrence in the query sequence. Especially, line1 indicates an *MUM*, since both occurrences are *unique supers*.

## *S1.2 When both non-embedded and embedded maximal matches are needed*

When both non-embedded and embedded maximal matches are needed, we have to turn on option *–ao* to include occurrences of *nested local* in the output file. The output file has a similar structure with the sample output in *S1.1*, except for its third column indicating four additional integers: the total number of occurrences in the reference sequence, the number of nested occurrences in the reference sequence, the total number of occurrences in the query sequence and the number of nested occurrences in the query sequence, separated by colons; from the fourth column on, positions of all occurrences (*super*, *non-nested local* and *nested local*) are listed.

```
1    S:40608 114:AAAAAAAAAAAAAAAAAAAAAAAAAAAAAAAAAAAAAAAATTAATACATGATTTGATTTTAAATATTAAATGACTTTCTTTTT
     ATTTTCCTTCTCTTCTCTCTCACATTCTTAA 1:0:1:0 1:808567429 2:786292654
2    S:13896708 110:AAAAAAAAAAAATTAGCCGGGCATGGTGGCGGGCGCCTGTAGTCCCAGCTACTCGGGAGGCTGAGGCAGGAGAATGGCGT
     GAACCCGGGAGGCGGAGCTTGCAGTGAGCC 2:0:1:0 1:1069760770 1:-2726413043 2:205357112
3    L:38942624 101:AAAAAAATACGAAAACCAGTCAGGCGTGGCGGCGTGCGCCTGCAATCGCAGGCACTCGGCAGGCTGAGGCAGGAGAATCA
     GGCAGGGAGGTTGCAGTGAGC 8:7:3:2 1:179419450 1:1828466975 1:2313972991 1:2349068155 1:-1274131725
     1:-1067359730 1:-983496574 1:-400874385 2:1304088511 2:1340474564 2:-489670408
4    L:806971 102:AAAAAAAAAAAAAAAAAAAAAAAAAGACTGTAGGAGCATCTGGTGGGAGGTGGTGGAGGGAGAACTGTGGGTTTGGAAGCTG
     CGCCCTCCCCCCAGCCATGC 2:2:3:2 1:2508659569 1:2534474142 2:2597025985 2:-2593711962 2:-2562502549
5    L:1004765 106:AAAAAAAAAAAAAAAAAAAAAAAAGAAGGAAGGAAGGGCCCAGAAGTCAGGAAGGAGCACGTGAGGAGGGTGTGTGGGAAG
     AATGGAGGTACTGAGGCAGGGTGCA 2:1:1:1 1:2499402921 1:-2504991767 2:2562500536
6    L:689564 110:AAAAAAAAAAAAAAAAAAAAAAAAAAGTCAGGAAACAACAGGTGCTGGAGAGGATGTGGAGAAATAGGAACACTTTTACACT
     GTTGGTGGGACTGTAAACTAGTTCAACC 4:0:1:0 1:-2897341951 1:-2290382390 1:-2216188336 1:-1822542408
     2:3166049008
7    L:14377644758 172:GTTGTTGAATTTTGTCAAAGGCCTTTTCTGCATCTATTGAGATAATCATGTGGTTTTTGTCTTTGGTTCTGTTTATA
     TGCTGGATTACGTTTATTGATTTTCATATGTTGAACCAGCCTTGCATCCCAGGGATGAAGCCCACTTGATCATGGTGGATAAGCTTTTTGATGTG
     3:2:3:2 1:1135681928 1:-916485511 1:-551977911 2:3096932611 2:-2709643878 2:-891973215
```

For super maxmers, the occurrences shown here are exactly the same to those shown in *S1.1*, and the inference of non-embedded maximal matches from these occurrences is also the same. But for local maxmers, due to the existence of *nested local*, when inferring non-embedded/embedded maximal matches from the list of occurrences, we need to check the contexts of each pair of occurrences to make sure they do form a maximal match. The current version of *SEQANALYSIS* does not provide contexts of occurrences in the output file—we have to refer to the original sequences in the input files, and check out the immediate left and right contexts of each occurrence. For example, for line 7 in the above example output, we attach contexts to each occurrence:

```
7    L:14377644758 172:GTTGTTGAATTTTGTCAAAGGCCTTTTCTGCATCTATTGAGATAATCATGTGGTTTTTGTCTTTGGTTCTGTTTATA
     TGCTGGATTACGTTTATTGATTTTCATATGTTGAACCAGCCTTGCATCCCAGGGATGAAGCCCACTTGATCATGGTGGATAAGCTTTTTGATGTG
     3:2:3:2 1:1135681928:C:C 1:-916485511:G:A 1:-551977911:G:C 2:3096932611:G:T 2:-2709643878:T:T
     2:-891973215:A:A
```

We compare the contexts of each occurrence in the reference sequence to those of each occurrence in the query sequence; only when a pair of occurrences have different contexts on both sides, does it form a maximal match. Occurrences of local maxmer in the reference sequence that form maximal matches with every occurrence in the query sequence are non-nested locals, and *vice versa*. We discriminate all occurrences in line 7 as

| **Compared sequences** | **Types of occurrence** | **Contexts** (left:right) | **positions of occurrences** |
|---|---|---|---|
| 1 (reference) | *non-nested* | C:C | 1135681928 |
| | *Nested* | G:A | -916485511 |
| | | G:C | -551977911 |
| 2 (query) | *non-nested* | T:T | -2709643878 |
| | *Nested* | G:T | 3096932611 |
| | | A:A | -891973215 |

According to table 4 in section 5.2 of the main text, these occurrences form the following maximal matches:

| **Maximal matches** | **Combination of occurrences** | **Occurrence pairs** |
|---|---|---|
| *Non-embedded* | *non-nested local + non-nested local* | 1135681928:C:C *vs* -2709643878:T:T |
| *Embedded* | *non-nested local + nested local* | 1135681928:C:C *vs* 3096932611:G:T<br>1135681928:C:C *vs* -891973215:A:A<br>-916485511:G:A *vs* -2709643878:T:T<br>-551977911:G:C *vs* -2709643878:T:T |
| | *nested local + nested local* | -551977911:G:C *vs* -891973215:A:A |

## ***Supplementary material 2:*** *Error estimation by the ratio of overlaps among non-embedded maximal matches for lineal ortholog identification*

Overlaps among non-embedded maximal matches (*nem* for short) enable us to estimate the false positive rate in our lineal ortholog identification in section 2.3, i.e., the proportion of collateral orthologs or paralogs (*cop* for short) among all non-embedded maximal matches. This conditional probability, $P(cop|nem)$ is the complement of $P(lo|nem)$ shown in figure 6, where *lo* stands for lineal orthologs. As we have mentioned in the main text, lineal orthology represents a one-to-one relationship of nucleotides; lineal orthologs by definition have no overlap—either "full overlap" or "partial overlap"—with one another: if two maximal matches overlap with each another, then in the overlapping region at least one of them doesn't represent a lineal ortholog. Supplementary figure 1 shows that misidentification of lineal orthologs often yields full or partial overlaps among non-embedded maximal matches as byproducts, and the number of *nem* entwined in such overlaps is often greater than the number of *cop* misidentified as *nem*. Consequently, the probability of overlaps among *nem* is expected to be higher than the probability of false positives in our lineal ortholog identification (equation 1); it therefore becomes possible to estimate an upper bound on $P(cop|nem)$ from the proportion of overlaps among *nem*, $P(overlap|nem)$. Supplementary figure 4 shows conditional probabilities from the numerical simulations in section 3.1 that a *nem* (a) is a *cop* given that it overlaps with other *nem*; and (b) overlaps with other *nem* given that it is a *cop*. Within the scope of our investigation in this paper, when the evolutionary distance between the compared genomes is not too far, probability (b) is always greater than probability (a): $P(overlap \mid nem \,\&\, cop) > P(cop \mid nem \,\&\, overlap)$ . According to Bayes' theorem,

$$\mathrm{P}(nem \,\&\, cop) = \frac{\mathrm{P}(cop \,\&\, nem \,\&\, overlap)}{\mathrm{P}(overlap \mid nem \,\&\, cop)}$$

$$= \frac{\mathrm{P}(cop \mid nem \,\&\, overlap) \times \mathrm{P}(nem \,\&\, overlap)}{\mathrm{P}(overlap \mid nem \,\&\, cop)} \quad .$$

Empirically, $\dfrac{\mathrm{P}(cop \mid nem \,\&\, overlap)}{\mathrm{P}(overlap \mid nem \,\&\, cop)} < 1$ (see supplementary figure 4) so that

$$P(nem \,\&\, cop) < P(nem \,\&\, overlap).$$

Dividing the above inequality by $P(nem)$ on both sides, we get

$$\boldsymbol{P(cop \mid nem) < P(overlap \mid nem)} \qquad \text{(Equation 1)}$$

That is to say, the probability that a *nem* represents *cop* can be expected to be smaller than the fraction of overlaps among all *nem*—the latter can be computed directly from the data; this offers a way to estimate an upper bound on the false positive errors in our lineal ortholog identification. For example, between orthologous chromosomes like human chromosome 1 and chimpanzee chromosome 1, among non-embedded maximal matches no shorter than 20 bases, the fraction of full or partial overlaps is about 24%, and the probability of misidentification can be expected to be lower. Therefore in natural genomes, especially in orthologous regions of closely-related genomes, a low rate of false positives can often be anticipated.

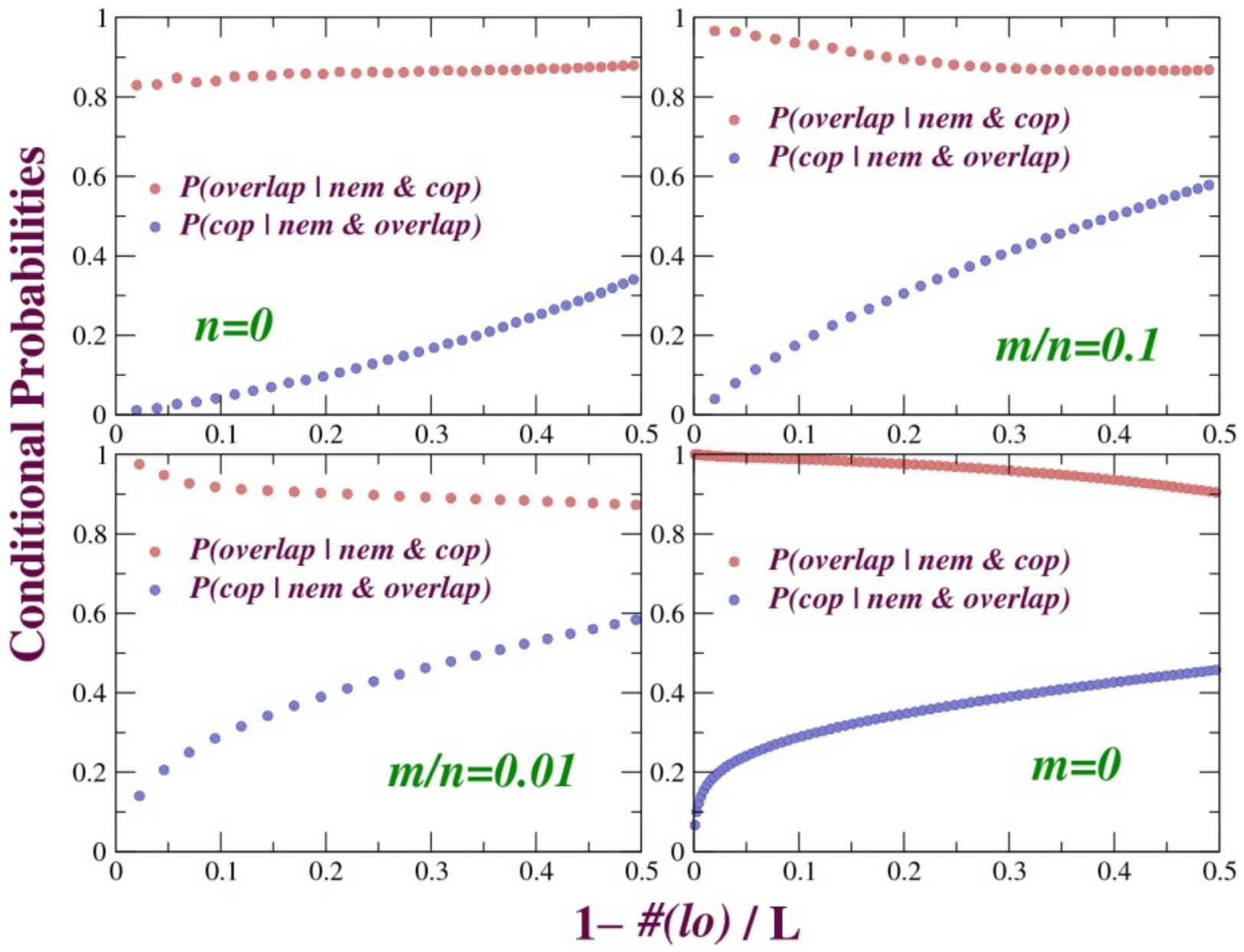

Supplementary figure 4. From the numerical simulations in section 3.1, conditional probabilities that (a) *P*(*cop* | *nem* & *overlap*): non-embedded maximal matches that overlap with other non-embedded maximal matches are collateral orthologs or paralogs; and that (b) *P*(*overlap* | *nem* & *cop*): collateral orthologs or paralogs misclassified as non-embedded maximal matches overlap with other non-embedded maximal matches; probabilities are counted for all maximal matches no shorter than 20 bases, and weighted by maximal match lengths.